\documentclass[sigconf]{acmart}

\setcopyright{acmcopyright}
\copyrightyear{2025}
\acmYear{2025}
\acmDOI{XXXXXXX.XXXXXXX}
 
\acmConference[IR-RAG Workshop at SIGIR '25]{Make sure to enter the correct
  conference title from your rights confirmation email}{July 17,
  2025}{Padua, Italy}
\graphicspath{{./images/}} 
\usepackage{bookmark}
\usepackage[frozencache=true,cachedir=minted-cache]{minted}
\usepackage{url}
\usepackage[show]{chato-notes}
\usepackage{tabularx}
\usepackage{multirow}
\usepackage{makecell}
\usepackage{calrsfs}
\usepackage{booktabs}
\usepackage{adjustbox}
\usepackage{caption}
\usepackage{subcaption}
\usepackage{tikz}
\usetikzlibrary{positioning}
\usepackage{mathtools}
\usepackage{amsmath}
\usepackage{soul}
\usepackage{tcolorbox}

\usepackage{pifont}

\newcommand{\numI}{$Iter$}

\usepackage{enumitem} 

\usepackage{caption}
\newcommand{\uls}{\begin{itemize}[leftmargin=*]}
\newcommand{\ule}{\end{itemize}}
\newcommand{\ols}{\begin{enumerate}[leftmargin=*]}
\newcommand{\ole}{\end{enumerate}}
\newcommand{\li}{\item}

\newcommand{\para}[1]{\paragraph{\textnormal{\textbf{#1}}}} 

\usepackage{multicol}
\usepackage{marginnote}

\DeclareMathAlphabet{\pazocal}{OMS}{zplm}{m}{n}
\DeclareMathAlphabet{\pazobfcal}{OMS}{cmsy}{b}{n}

\usepackage{tikz}

\newcommand{\cm}[1]{\textcolor{black}{#1}}
\newcommand{\jy}[1]{\textcolor{black}{#1}}
\newcommand{\newcontent}[1]{\textcolor{black}{#1}}

\begin{document}



\title[Am I on the Right Track?]{Am I on the Right Track? What Can Predicted Query Performance Tell Us about the Search Behaviour of Agentic~RAG}

\author{Fangzheng Tian}
\email{f.tian.1@research.gla.ac.uk}
\affiliation{%
  \institution{University of Glasgow}
  \city{Glasgow}
  \country{United Kingdom}
}

\author{Jinyuan Fang}
\email{j.fang.2@research.gla.ac.uk}
\affiliation{%
  \institution{University of Glasgow}
  \city{Glasgow}
  \country{United Kingdom}
}

\author{Debasis Ganguly}
\email{First.Last@glasgow.ac.uk}
\affiliation{%
  \institution{University of Glasgow}
  \city{Glasgow}
  \country{United Kingdom}
}

\author{Zaiqiao Meng}
\email{First.Last@glasgow.ac.uk}
\affiliation{%
  \institution{University of Glasgow}
  \city{Glasgow}
  \country{United Kingdom}
}

\author{Craig Macdonald}
\email{First.Last@glasgow.ac.uk}
\affiliation{%
  \institution{University of Glasgow}
  \city{Glasgow}
  \country{United Kingdom}
}


\renewcommand{\shortauthors}{Tian et al.}


\begin{abstract}

\looseness -1 Agentic Retrieval-Augmented Generation (RAG) is a new paradigm where the reasoning model decides when to invoke a retriever (as a ``tool'') when answering a question. 
This paradigm, exemplified by recent research works such as Search-R1, enables the model to decide when to search and obtain external information.
However, the queries generated by such Agentic RAG models and the role of the retriever in obtaining high-quality answers remain understudied. To this end, this initial study examines the applicability of query performance prediction (QPP) within the recent Agentic RAG models Search-R1 and R1-Searcher. We find that applying effective retrievers can achieve higher answer quality within a shorter reasoning process. Moreover, the QPP estimates of the generated queries, used as an approximation of their retrieval quality, are positively correlated with the quality of the final answer. Ultimately, our work is a step towards adaptive retrieval within Agentic RAG, where QPP is used to inform the model if the retrieved results are likely to be useful.
\end{abstract}



\maketitle

\section{Introduction}

Retrieval-Augmented Generation (RAG) is a technique that incorporates retrieved external knowledge into the generative process of a Large Language Model (LLM)~\cite{ragReview}. By doing so, RAG alleviates the hallucination problem often observed in LLMs~\cite{originOfHallucinations} and compensates for their limitations in accessing up-to-date or domain-specific knowledge~\cite{ragInKnowledgeIntensiveNLP}, which is essential for fulfilling downstream tasks such as Question Answering (QA). \cm{The basic {\em standard} form} of the RAG pipeline involves a single retrieval stage, based solely on an input question. However, standard RAG has limitations in its flexibility and adaptability, as it applies a uniform retrieval strategy to all input questions, regardless of whether the LLM already possesses sufficient internal knowledge to answer them. In addition, it is not well-suited for tasks that require complex or multi-step information gathering~\cite{FLARE}. As a result, the potential benefits of integrating retrieval and generation are not fully realised, which can lead to suboptimal performance on complex downstream tasks.

To address the limitation of single retrieval in standard RAG, an alternative approach is to delegate the LLM itself to determine when to invoke searches and what queries to search for. This paradigm is termed as {\em iterative} or {\em Agentic RAG}.
Instead of relying on a static, single-step retrieval, the LLM in Agentic RAG can iteratively formulate and issue queries, and ingest the results of the search before continuing its reasoning process. \cm{This allows the reasoning process to adapt in light of the retrieved information}. Agentic RAG models such as Search-R1~\cite{searchR1} and R1-Searcher~\cite{r1Searcher} have shown improved performance in question answering tasks, such as Natural Questions (NQ)~\cite{NQ} and HotpotQA~\cite{hotpotQA}.

In the workflow of Agentic RAG, an LLM begins answering a question by reasoning. During the reasoning process, if external knowledge is needed, the LLM can automatically generate intermediate queries that interrupt generation and trigger the search engine (acting as an agent). The top-ranked retrieved documents are then appended to the generated tokens, serving as the context for reasoning in the next iteration. These reasoning-retrieval iterations continue until the LLM decides to yield the final answer after reasoning or reaching the output token limit. While existing Agentic RAG models have shown effective empirical results \newcontent{in question-answering tasks}~\cite{searchR1,r1Searcher}, there remains a limited understanding of the properties of the generated queries and their retrieval results, as well as how the retrieval component contributes to enhancing answer quality. To further enhance the performance of Agentic RAG by improving the reasoning-retrieval process, it is important to analyse the generated queries \cm{and their likelihood of retrieving relevant information}. \cm{This motivates us to study Agentic RAG's retrieval component and its downstream effects on answer quality.}

While the role of the retrieval component in standard RAG systems has been extensively studied, its role within Agentic RAG remains under-explored. In standard RAG, factors such as retriever effectiveness, the cutoff of top-retrieved documents and the order of documents when they are appended into RAG prompts can all influence the answer quality. However, Agentic RAG systems introduce additional complexity by involving multiple retrievals throughout the reasoning process. Therefore, in Agentic RAG, answer quality may be affected by the cumulative influence of all the retrievals conducted during the reasoning iterations, presenting new challenges for analysing it.

\looseness -1 A significant challenge in evaluating the impact of retrieval quality within the Agentic RAG framework is the absence of ground truth relevance judgements for the intermediate generated queries in each reasoning-retrieval iteration. To tackle this challenge, we resort to Query Performance Prediction (QPP) as a possible estimator of the retrieval quality for generated queries. QPP methods aim to predict the effectiveness of the retrieval for a query without human-annotated relevance judgements, typically by analysing features of the query and the retrieved documents \cite{NQC,aPairRatio,WRIG}. Given that both the generated queries and their corresponding retrieval results are available within the Agentic RAG (implemented by Py-Terrier-RAG~\cite{macdonald25pyterrier_rag}), QPP offers a viable approach for approximating retrieval quality and investigating the role of retrieval in the system.

To this end, in this paper, we conduct experiments examining answer quality for two recent Agentic RAG models using three different retrieval models. For each retriever, we study how estimated query performance, estimated by several unsupervised QPP models, varies throughout the iterations in the reasoning process. Furthermore, we investigate how the estimated query performance correlates with answer quality. 

In summary, this paper contributes a first study of the utility of QPP in the querying behaviour of two typical agentic RAG models, Search-R1~\cite{searchR1} and R1-Searcher~\cite{r1Searcher}. Moreover, we investigate the impact of retriever effectiveness on the iterative reasoning-retrieval process in Agentic RAG. We conduct experiments on the 3610 test queries of the Natural Questions (NQ) QA dataset~\cite{NQ}. Based on our experiment results, we observe that (1) more effective retrieval models can reduce the length of Agentic RAG's reasoning process and enhance the answer quality; (2) QPP estimation about the generated queries is a potential signal for predicting the quality of the final answers.

The remainder of the paper is organised as follows.
In the next section, we review related work about RAG and the role of the retrieval component in it. We then provide a systematic description of Agentic RAG system in Section~\ref{s:system_description}. In Section~\ref{s:settings} we state our research questions and introduce the setting of Agentic RAG pipelines and the experimented QPP methods. According to the experiment results, we answer the research questions in Section~\ref{s:results} and conclude the paper with perspectives on applying QPP in facilitating Agentic RAG in Section~\ref{s:conclusion}. \newcontent{Appendix~\ref{sec:appendix} demonstrates a case study on two outputs - one with a correct answer and one with a wrong answer - to investigate the topical changes in the intermediate queries.}

\begin{figure}[tb]
\centering
\includegraphics[width=0.99\columnwidth]{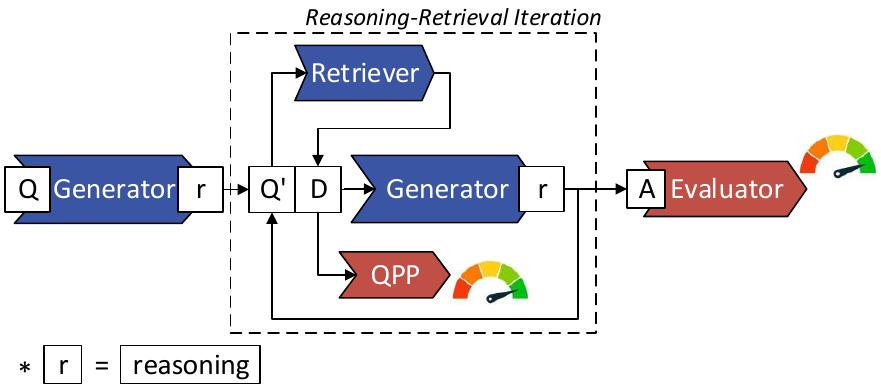}
\caption{
A conceptual overview of a search-enhanced reasoner based RAG, which we simply refer to as ``Agentic RAG''. Given an input question $Q$, the model performs initial reasoning $r$, following which it iteratively generates queries $Q'$. For each such generated query, the retriever component incorporates the top-ranked documents into the ongoing reasoning process. This process continues until the model produces a final answer $A$ after \numI{} iterations (the number of iterations required depends on the input question). A post-retrieval query performance prediction (QPP) model is applied to the intermediate retrieval results. An evaluator then assesses the quality of the generated answer. In this paper, we seek to find correlations between the QPP estimates and the answer quality.}
\label{fig:flowchart}
\end{figure}

\begin{figure*}[tb]
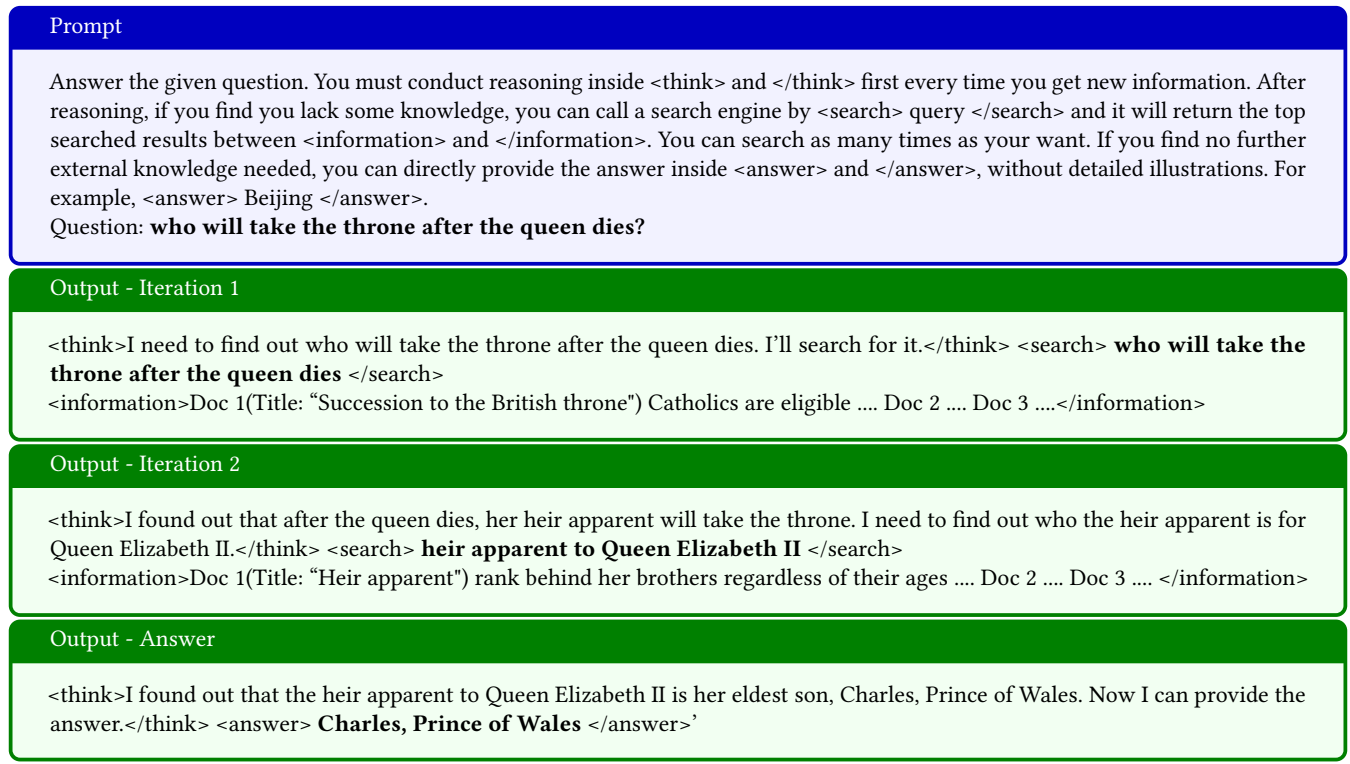

\begin{subfigure}{\textwidth}
\begin{tcolorbox}[colback=blue!5!white, colframe=blue!75!black, title=Prompt]

Answer the given question. You must conduct reasoning inside <think> and </think> first every time you get new information. After reasoning, if you find you lack some knowledge, you can call a search engine by <search> query </search> and it will return the top searched results between <information> and </information>. You can search as many times as your want. If you find no further external knowledge needed, you can directly provide the answer inside <answer> and </answer>, without detailed illustrations. For example, <answer> Beijing </answer>.

Question: \textbf{who will take the throne after the queen dies?}

\end{tcolorbox}
\end{subfigure}
\hfill
\begin{subfigure}{\textwidth}
\begin{tcolorbox}[colback=green!5!white, colframe=green!50!black, title=Output - Iteration 1]

<think>I need to find out who will take the throne after the queen dies. I'll search for it.</think> <search> \textbf{who will take the throne after the queen dies} </search>

<information>Doc 1(Title: ``Succession to the British throne") Catholics are eligible .... Doc 2 .... Doc 3 ....</information>
\end{tcolorbox}
\end{subfigure}
\hfill
\begin{subfigure}{\textwidth}
\begin{tcolorbox}[colback=green!5!white, colframe=green!50!black, title=Output - Iteration 2]

<think>I found out that after the queen dies, her heir apparent will take the throne. I need to find out who the heir apparent is for Queen Elizabeth II.</think> <search> \textbf{heir apparent to Queen Elizabeth II} </search>

<information>Doc 1(Title: ``Heir apparent") rank behind her brothers regardless of their ages .... Doc 2 .... Doc 3 .... </information>
\end{tcolorbox}
\end{subfigure}
\hfill
\begin{subfigure}{\textwidth}
\begin{tcolorbox}[colback=green!5!white, colframe=green!50!black, title=Output - Answer] 

<think>I found out that the heir apparent to Queen Elizabeth II is her eldest son, Charles, Prince of Wales. Now I can provide the answer.</think> <answer> \textbf{Charles, Prince of Wales} </answer>'
\end{tcolorbox}
\end{subfigure}
\caption{\looseness -1 Example prompt for Search-R1 and its output over two reasoning-retrieval iterations for a query from the Natural Questions dataset. In each iteration, the top three documents retrieved by E5 are used to inform the reasoning process. In this example, the generator is tasked with identifying the successor to the late Queen. Search-R1 performs two iterations with generated queries before producing the correct answer.}
\label{fig:output_example}

\end{figure*}

\section{Related Work}\label{s:review}
We position our work within the existing literature, in terms of retrievers in RAG (Section~2.1), iterative RAG models (Section~2.2) and query performance prediction (Section~2.3).

\subsection{The Role of Retriever in RAG}
\looseness -1 \jy{RAG models enhance an LLM's generative process by providing access to external knowledge~\cite{ragInKnowledgeIntensiveNLP}, which helps mitigate hallucinations~\cite{detectHallucinations} and improve overall performance.} In RAG, particularly for the decoder-only LLMs~\cite{llama3,fewShotLearnerLLM,qwen25}, retrieved documents are directly concatenated with the input question~\cite{ragReview} as part of the prompt. RAG has been \cm{applied to a number of downstream tasks}, including QA~\cite{FiD,ragInKnowledgeIntensiveNLP}, fact verification~\cite{selfRAG,ATLAS} and summarisation~\cite{ragCodeGenAndSummarisation}.

Unlike traditional information systems, where users consume the retrieval results, RAG systems pass the retrieval results directly to a generator (i.e. an LLM). 
\jy{In RAG systems, it is the LLM, rather than the user, that consumes the retrieved documents to generate the final response. Traditional IR metrics, such as nDCG, may not precisely measure the usefulness of retrieved documents in this context, due to the difference between human-defined relevance and the way LLMs interpret and leverage information~\cite{evalGenIR}. Therefore, the quality of retrieved documents in RAG should be evaluated not by topical relevance but rather by their impact on the downstream task performance~\cite{ragRelevanceLabel,evalGenIR,RelevanceAndUtility}. To better understand this relationship, recent studies have examined how different characteristics of the retrieved documents affect RAG performance, such as the number of documents~\cite{RelevanceAndUtility}, the position of relevant documents~\cite{lostInTheMiddle} and the relative position of the query~\cite{powerOfNoise}.}
Despite these findings, existing studies about the role of retrieval in RAG focus on single-retrieval RAG pipelines. There remains a \textbf{gap in understanding how retrieval influences performance in RAG pipelines that involve multiple retrievals.}


\subsection{Iterative/Agentic RAG}
\looseness -1 
\jy{In standard RAG models,} external knowledge is retrieved and provided to the LLM only once~\cite{ragInKnowledgeIntensiveNLP}. However, a single retrieval might not be enough to answer some difficult queries~\cite{covRAG,efficientRAG}. To address this issue, recent studies, such as IRCoT~\cite{trivedi2023interleaving}, KiRAG~\cite{fang2025kirag} and Search-O1~\cite{searchO1}, have explored RAG frameworks that incorporate multiple (or iterative) retrievals. The number of retrievals can be determined dynamically using techniques such as question decomposition~\cite{FLARE} or by iteratively detecting uncertain tokens in the generated answer~\cite{RAT}. The queries executed in these dynamic retrievals are often refined versions of the original query~\cite{reasoningPlaningRAG} or are generated to resolve ambiguities in the current answer~\cite{FLARE}.

More recent RAG frameworks integrate dynamic retrieval directly into the LLM's reasoning process~\cite{searchO1,r1Searcher,searchR1}, letting LLM itself autonomously conduct searches. This is a new paradigm known as Agentic RAG~\cite{agenticRAGSurvey}. Search-O1 achieves this integration by prompt engineering~\cite{searchO1}, whereas more advanced models such as Search-R1~\cite{searchR1} and R1-Searcher~\cite{r1Searcher} employ reinforcement learning to train the LLM to autonomously invoke retrievals during the reasoning process. These methods have demonstrated improved performance on QA benchmarks~\cite{NQ,hotpotQA}. In this paper, \textbf{we investigate the reasoning-retrieval iterations in RL-trained Agentic RAG models, focusing on the role of the retriever and its impact on answer quality.}

\subsection{Query Performance Prediction}
\looseness -1 Query performance prediction (QPP) is a technique used to estimate the quality of retrieval results for an input query without human-assessed relevance judgements. QPP methods are generally categorised into \textbf{pre-retrieval} and \textbf{post-retrieval}, based on whether the prediction is made before or after the retrieval process.

Pre-retrieval QPP methods solely rely on the features of the query itself~\cite{preRetrievalSurvey} -- such as inverse document frequency (IDF)~\cite{spire2004}, query length -- to infer the query's specificity and predict the retrieval performance. Because they do not consider retrieval results, their predictions are independent of the retriever used and are generally less accurate~\cite{qppFromAdHocToConvSearch}.

\looseness -1 In contrast, post-retrieval QPP methods can use the features of the retrieval results, including retrieval score~\cite{NQC,SMV,UEF,Clarity}, dense embeddings~\cite{denseQPP,aPairRatio} and the document texts~\cite{bertQPP}. Among numerous post-retrieval QPP methods, supervised methods directly train models to predict specific target IR metrics~\cite{bertQPP,qppBERT-PL}. Unsupervised post-retrieval QPP models analyse the retrieval quality by analysing certain properties of the retrieval result, such as the distinctiveness of the top-retrieved document~\cite{Clarity,NQC}, semantic concentration~\cite{denseQPP}, coherence~\cite{coherenceQPP,aPairRatio}, etc. \textbf{However, QPP methods have not been applied in examining the query generation process in Agentic RAG.} Given the lack of ground truth for evaluating the queries generated during reasoning-retrieval iterations, QPP provides a promising technique for investigating the role of retrieval in Agentic RAG systems.

\section{Agentic RAG}\label{s:system_description}


In this section, we describe how we incorporate a query performance prediction (QPP) component in a search-enhanced reasoning-based RAG workflow, which, for simplicity, we simply refer to as \textbf{Agentic RAG}, as the LLM uses the retriever as an agent.

\subsection{Overview}~\label{ss:system_overview}

Figure~\ref{fig:flowchart} shows the workflow of a Search-Enhanced Reasoning based RAG, which we simply refer to as ``Agentic RAG'' throughout the paper. In this system, an LLM-decoder, serving as the \textbf{generator}, automatically drives a \textbf{retriever} for searching information. Given an input question $Q$, the generator performs reasoning towards deriving an answer. During the reasoning process, if the generator perceives that additional external information is necessary, it generates a query $Q'$, enclosed within the tags {\tt <search>} and {\tt </search>}.\footnote{For clarity, in this paper, we always refer to the user's input as a question, and the search queries generated by the model as queries.} The retriever then executes this query to retrieve potentially relevant documents, which are incorporated into the ongoing reasoning process. This may prompt the generator to initiate further retrievals. Each cycle of this `reasoning' and `retrieval' constitutes an \textit{iteration} in agentic RAG. This iterative process continues until the generator decides to yield the final answer $A$. The total number of \newcontent{reasoning-retrieval} iterations undertaken to reach the final answer for the input query is termed as \textbf{length of the reasoning process}, denoted as $Iter$ in this paper.

To exemplify the reasoning process of agentic RAG, Figure~\ref{fig:output_example} presents a sample prompt and output from Search-R1, which is a commonly used search-enhanced reasoning based RAG model~\cite{searchR1}.
The top-part of the figure shows the prompt that explicitly instructs the model to invoke searches when more knowledge is required for answering the input question.
In this example, the generator invokes the retriever twice with progressively refined queries before yielding the final answer. In this example, the answer generated after two steps ($Iter$=2) correctly answers the input question about the successor of the late Queen.

\subsection{Retrieval}~\label{ss:overview_retrieval}

When configuring a RAG model, various retrieval pipelines can be integrated with the generator. They can range from an efficient sparse retriever, such as BM25, to more complex retrieval-re-ranking pipelines. Using PyTerrier-style notation, we can denote any retrieval pipeline for RAG. For instance, a MonoT5 cross-encoder re-ranking the output of BM25 as: \texttt{BM25$\mbox{ }\%\mbox{ }k$ $\gg$ MonoT5}\footnote{In 
PyTerrier~\cite{pyterrier}, $a$$\gg$$b$ denotes a pipeline composition operation: the output of $a$ is passed to $b$; $\%k$ denotes applying a rank cutoff of $k$ per query.}. Alternatively, a dense retrieval pipeline may be formulated with a query-encoder and an existing vector database of pre-encoded passage vectors (e.g. E5 $\gg$ VectorIndex).

This flexibility in configuring retrieval pipeline allows us to easily instantiate Agentic RAG models with various retrievers\footnote{An example of RAG pipeline with BM25 $\gg$ MonoT5 as the retriever can be found here: \url{https://github.com/terrierteam/pyterrier_rag/tree/main/examples}.}, which can result in different downstream RAG performances. Intuitively, the top-$k$ documents (where $k$ is typically small in RAG) retrieved by more effective retrievers are more likely to contain relevant information. Such documents can better complement the LLM’s internal knowledge, potentially reducing the reasoning-retrieval iterations required to generate an accurate answer. However, since retrieval is only an intermediate step in the Agentic RAG, the degree to which the retrieval pipeline impacts answer quality remains an open question—one that we investigate in our research.


\subsection{QPP of Generated Queries}

To investigate the influence of the retrieval component in Agentic RAG models, we need to first estimate the quality of the retrieved documents. In this context, retrieval quality refers to how relevant the retrieved documents are to the generated sub-query, and how useful these documents will be to `guide' the generator towards a high-quality (relevant and factually correct) answer.

Since the retrieval component receives queries that are dynamically generated by the LLM's reasoning process, there is no relevance ground truth for explicitly evaluating retrieval quality. To address this problem, in this initial research, we investigate whether a Query Performance Predictor (QPP) estimate may be substituted as a reasonable approximation for the relevance or the usefulness of these documents retrieved during the reasoning phase.

In this work, we focus on post-retrieval QPP models, which estimate retrieval quality based on properties of the retrieved documents, such as their score distribution or embedded representations, instead of pre-retrieval ones, where the prediction depends only on the input query characteristics. A post-retrieval QPP model is a more appropriate choice for this task of estimating the usefulness of the retrieved documents in guiding the reasoning process towards the correct answer.

A QPP model can be integrated within the Agentic RAG pipeline to dynamically estimate the quality of retrievals in each reasoning-retrieval iteration. This component is depicted as the red box connected to the retrieval result in Figure~\ref{fig:flowchart}.

\looseness -1 Inclusion of a QPP model into the agentic RAG workflow is a step towards better understanding the quality of the intermediate queries generated during the reasoning-retrieval iterations. 
This is also a step towards developing an adaptive agentic RAG workflow that is capable of dynamically deciding to leverage the top-retrieved documents for modifying the reasoning process only if their estimated performance is high. In this initial work, we focus only on the analysis of the QPP estimates and its relation with with the downstream answer quality, leaving the exploration of an adaptive QPP based workflow for agentic RAG as future work.

Finally, it is worth mentioning that the pragmatic applicability of QPP methods depends on the retrieval model type used in RAG. 
For instance, it is not a pragmatic choice to apply dense QPP approaches, such as ~\cite{denseQPP,aPairRatio}, to retrievers using sparse representations (e.g., BM25). That is because if document embeddings are available, we may as well use them to improve the retriever itself, e.g., bi-encoder retriever, instead of solely utilising them for QPP~\cite{aPairRatio}. 


\section{Experimental Setup}\label{s:settings}

We now elicit our research questions (Section~\ref{ssec:rq}), our experimental setup in terms of retrievers and agentic rag models (Section~\ref{ssec:pipelines} and query performance predicts (Section~\ref{ss:qpp_method_settings}).

\subsection{Research Questions}\label{ssec:rq}


\looseness -1 In this work, we focus on the iterative query generation process in two agentic RAG models, namely Search-R1~\cite{searchR1} and R1-Sear\-ch\-er~\cite{r1Searcher}, and specifically the quality of queries generated at each iteration. Indeed, if the effectiveness of the retriever used by the RAG system changes, the number of iterations required to reach a final answer may change. During their iterative reasoning–retrieval processes, the quality of the generated queries may shift -- becoming more specific or more generic -- which, in turn, can influence the final answer quality. Since human-assessed relevance judgments are not available for these dynamically generated queries, we estimate their effectiveness using query performance predictors. Based on this, we formulate our research questions as follows:

\uls
\li \textbf{RQ-1}: What is the effect of different retriever configurations on answer quality and number of iterations?
\li \textbf{RQ-2}: How does the estimated performance of generated queries vary as the number of iterations increases?
\li \textbf{RQ-3}: How does the estimated performance of generated queries correlate with the quality of the generated answers?
\ule

\subsection{Agentic RAG Pipeline Configurations}\label{ssec:pipelines}

\para{Query set} We use the Natural Questions (NQ) test set~\cite{NQ}, which contains a total of 3,610 questions. Each question is accompanied by gold (ground-truth) answers, as found from Wikipedia by manual annotation. These ground-truth answers are then used to evaluate the quality of the answers generated by Agentic RAG models.
Following common practices, we use F1 and Exact Match \% (EM) to measure answer quality~\cite{evaluatingQAEval}.

\para{Document corpus} A snapshot of English Wikipedia from 2018, as used with the NQ dataset.

\para{Generator} We experiment with two recent Agentic RAG models trained with Reinforcement Learning (RL) from QWEN2.5-7B~\cite{qwen25}:
\uls
\li \textbf{Search-R1}~\cite{searchR1}: This model is trained by single-stage outcome-based RL on NQ and HotpotQA training sets~\cite{hotpotQA}\footnote{\url{https://huggingface.co/PeterJinGo/SearchR1-nq_hotpotqa_train-qwen2.5-7b-em-ppo}}.
\li \textbf{R1-Searcher}~\cite{r1Searcher}: This model is trained via two-stage outcome-based RL on HotpotQA and 2WikiMultiHopQA~\cite{2WikiMultihopQA}. Since NQ was not used in the training of R1-Searcher model, our experiments for the R1-Searcher correspond to that of an out-of-domain evaluation. 
\ule
Both Search-R1 and R1-Searcher have demonstrated improved answer quality over standard RAG formulations and other recent iterative RAG formulations such as IRCoT~\cite{trivedi2023interleaving}.

\para{Retriever} We address both sparse/lexical retrieval, as well as neural re-ranking and dense retrieval:
    \ols
    \li \textbf{BM25}: A classic lexical retrieval model; 
    \li \textbf{MonoT5}: A retrieval-re-ranking pipeline with cross-encoder BM25$\gg$MonoT5. The top-20 documents retrieved by BM25 are re-ranked by MonoT5~\cite{monot5}. 
    \li \textbf{E5}: The bi-encoder model E5~\cite{wang2022text} for general purpose retrieval. It is commonly used as the retriever in RAG pipelines, including in the original papers for Search-R1 and R1-Searcher.
    \ole


The top-3 documents retrieved by each retrieval model are incorporated into the reasoning process. These models span a range of retrieval strategies with varying effectiveness. In general, re-ranking BM25 results with MonoT5 improves relevance, as the cross-encoder can better model semantic similarity. However, the utility of MonoT5 may be hindered by the limited recall offered by BM25 (relevant documents not retrieved in the top 20 by BM25 cannot be retrieved by MonoT5). As a semantic dense retrieval model, E5 may retrieve passages that do not match any of the terms in the queries - therefore, we expect that E5 should be more effective than MonoT5 (and BM25).




All of the above RAG and retrieval pipelines are implemented using PyTerrier~\cite{pyterrier}\footnote{\url{https://github.com/terrier-org/pyterrier}} and PyTerrier-RAG~\cite{macdonald25pyterrier_rag}\footnote{\url{https://github.com/terrierteam/pyterrier_rag}}.

\subsection{Query Performance Predictors}\label{ss:qpp_method_settings}

In our experiments, we apply QPP methods on the generated queries to estimate how relevant the retrieval results are to those generated queries, which may be related to the usefulness of those queries in answering the original question. Specifically, we test:

\para{Predictors based on Retrieval Scores} We experiment with
    \uls
    \li \textbf{NQC}~\cite{NQC}: NQC is the classic QPP method based on the standard deviation of retrieval scores for retrieved documents. It is applicable to all three retrieval configurations. We use the retrieval scores of the top-100 retrieved documents in NQC, which is a reasonable setting to acquire accurate QPP estimates~\cite{aPairRatio}.
    \li \textbf{Max(Score)}~\cite{aPairRatio}: For the neural retrievers, in which the retrieval scores can more accurately reflect the relevance between a query-document pair, there is a high likelihood that the score of the top-ranked document likely acts as an effective estimator of query performance. This predictor, denoted as \textbf{Max(Score)}~\cite{aPairRatio}, reflects the estimated upper bound of the relevance of the retrieved context, which is likely to be reliable due to a small cutoff on the retrieved lists, as usually is the case in Agentic RAG.
    \ule

\para{Embedding-based Predictors} Additionally, for the bi-encoder model (E5), which uses dense embeddings of the retrieved documents, we test two recent embedding-based QPP models:
    \uls
    \li \textbf{A-Pair-Ratio}~\cite{aPairRatio}: A coherence-based predictor that calculates the ratio between the pairwise coherences of the top-ranked and last-ranked documents in the retrieval result. In particular, we compute this ratio among the first five and the last five documents in the top-50 retrieved documents, as described in \cite{aPairRatio}.
    \li \textbf{Dense-QPP}~\cite{denseQPP}: A recent geometric-based predictor that calculates the volume of the minimum hypercube encompassing the embeddings of the top-$k$ retrieved documents and the query. We calculate the hypercube based on the top 3 documents, as per our RAG configuration.
    \ule

Finally, considering that there can be multiple intermediate generated queries for answering an input question, when evaluating the correlation between QPP estimates and the quality of generated answers, in this initial study, we only consider the QPP estimates for the first generated query.

\section{Results}\label{s:results}
We now report findings for each of our research questions.

\begin{table}[tb]
\centering
\caption{Average quality (measured by EM \& F1) of the answers generated by Search-R1 and R1-Searcher on the Natural Questions dataset, using various retrievers. We also report the average number of reasoning/querying iterations (denoted \numI{}).}\label{table:system_summarisation}
\begin{adjustbox}{width=\columnwidth}
\begin{tabular}{@{}l rrr rrr @{}}
\toprule
\small
Model & \multicolumn{3}{c}{Search-R1} & \multicolumn{3}{c}{R1-Searcher}  \\
\cmidrule(r){2-4} \cmidrule(r){5-7} 
 Retriever & BM25 & MonoT5 & E5 & BM25 & MonoT5 & E5 \\
\midrule 

 EM & 0.3391 & 0.3873 & 0.4838 & 0.2089 & 0.2307 & 0.3075  \\
\cmidrule(r){2-4} \cmidrule(r){5-7} 
  F1  & 0.4185 & 0.4709 & 0.5687 & 0.2586 & 0.2865 & 0.3736  \\
\midrule 
 \numI{} & 2.52 & 2.19 & 2.00 & 2.20 & 2.03 & 1.99  \\

\bottomrule
\end{tabular}
\end{adjustbox}
\end{table}

\begin{table}[tb]
\centering
\caption{Spearman's $\rho$ correlation between number of iteration (\numI{})
and answer quality (F1). 
}\label{table:iteration_length_correlation}
\begin{adjustbox}{width=\columnwidth}
\begin{tabular}{@{}l rrr rrr @{}}
\toprule
\small
Model & \multicolumn{3}{c}{Search-R1} & \multicolumn{3}{c}{R1-Searcher}  \\
\cmidrule(r){2-4} \cmidrule(r){5-7} 
 Retriever & BM25 & MonoT5 & E5 & BM25 & MonoT5 & E5 \\
\midrule 
 $\rho$(\numI{}, F1) & -0.2803 & -0.2991 & -0.3168 & -0.1872 & -0.1490 & -0.1110  \\
\bottomrule
\end{tabular}
\end{adjustbox}
\end{table}

\begin{figure*}[tb]
\centering
\begin{subfigure}{0.48\textwidth}
\includegraphics[width=\textwidth]{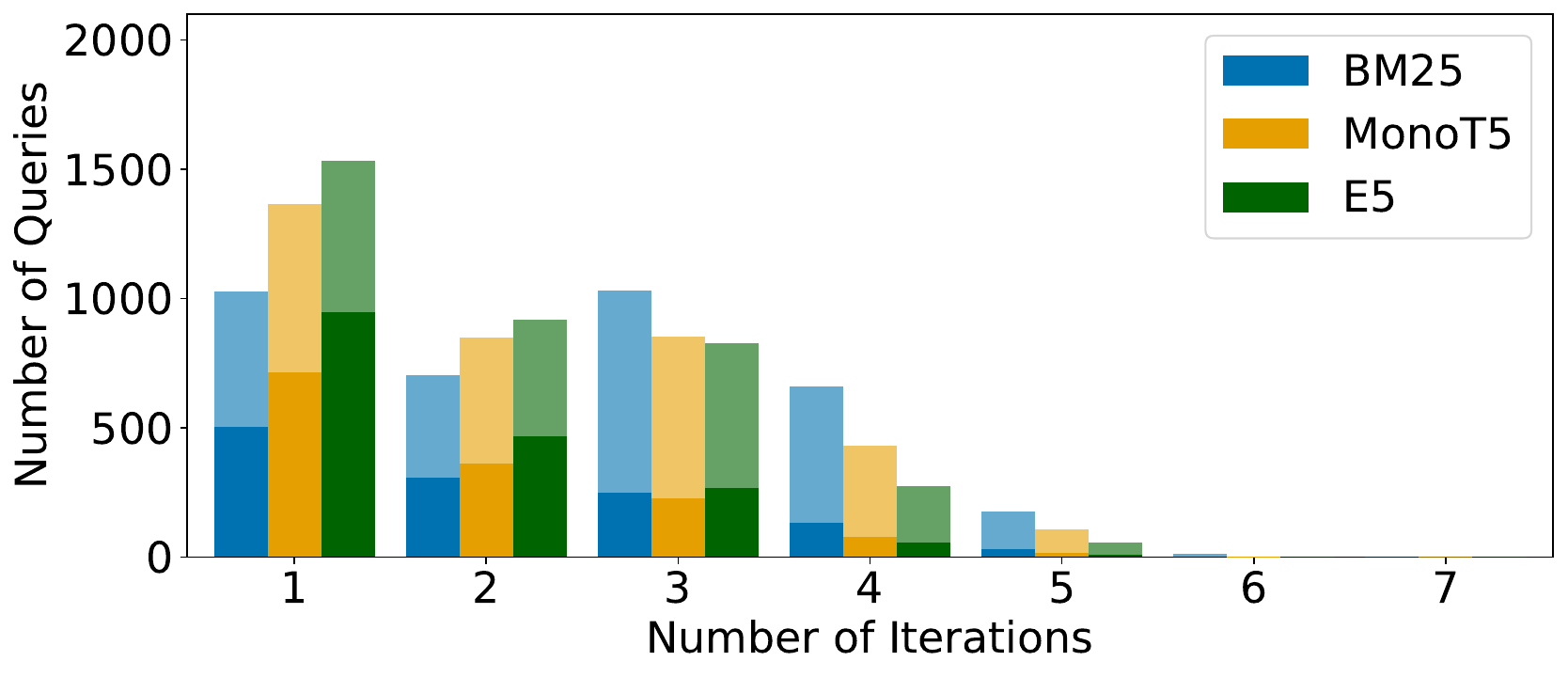}
\caption{Search-R1.}
\end{subfigure}
\hfill
\begin{subfigure}{0.48\textwidth}
\includegraphics[width=\textwidth]{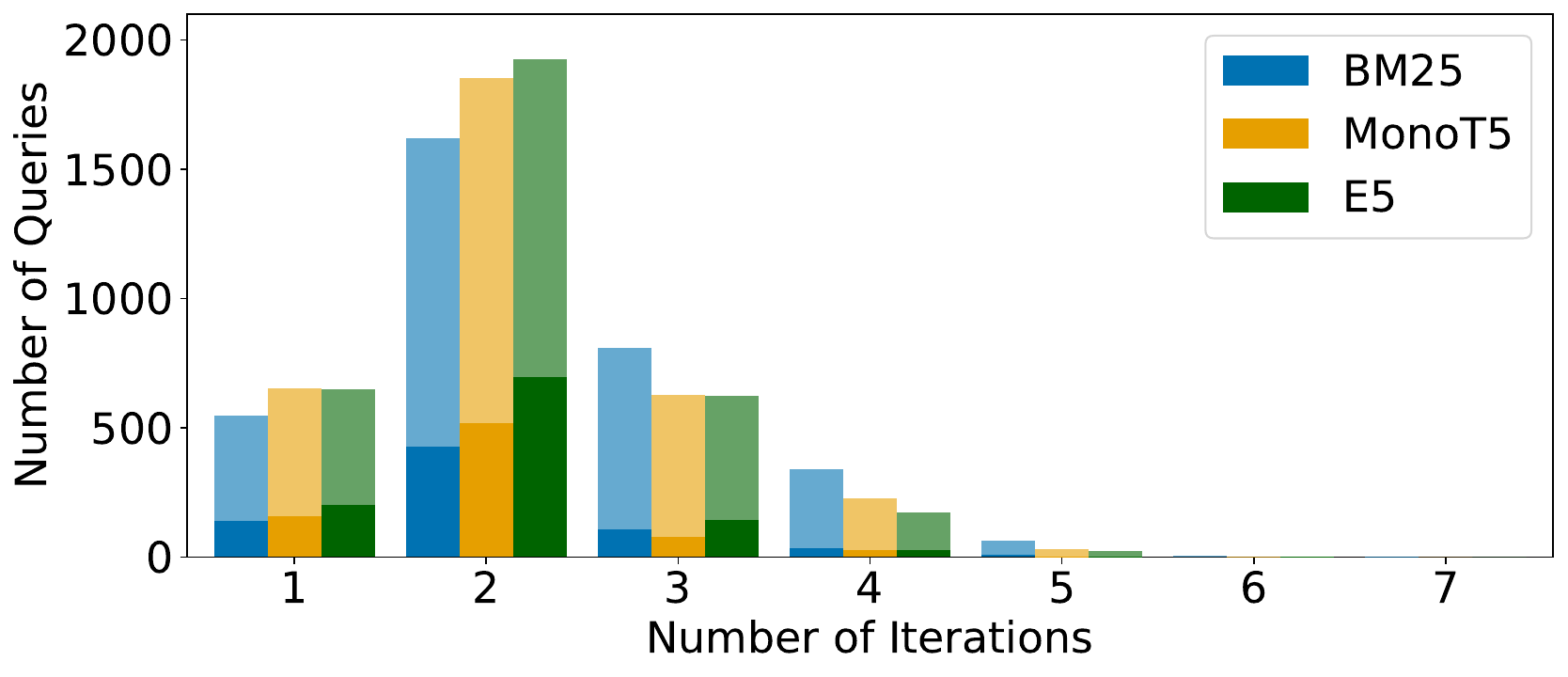}
\caption{R1-Searcher.}
\end{subfigure}
\vspace{-1em}
\caption{Distribution of the reasoning lengths (number of iterations) across 3610 NQ dataset questions for (a) Search-R1 and (b) R1-Searcher agentic RAG models. For both these models, we show the number of iterations for the three retrieval models experimented with. In each bar, the height of the part with a darker colour is the number of queries with a correct answer (EM=1). Notably, BM25 exhibits a longer tail with a higher number of reasoning iterations compared to that of the neural retrieval models, MonoT5 and E5.}
\label{fig:distribution}
\end{figure*}

\begin{figure}[tb]
\centering
\includegraphics[width=85mm]{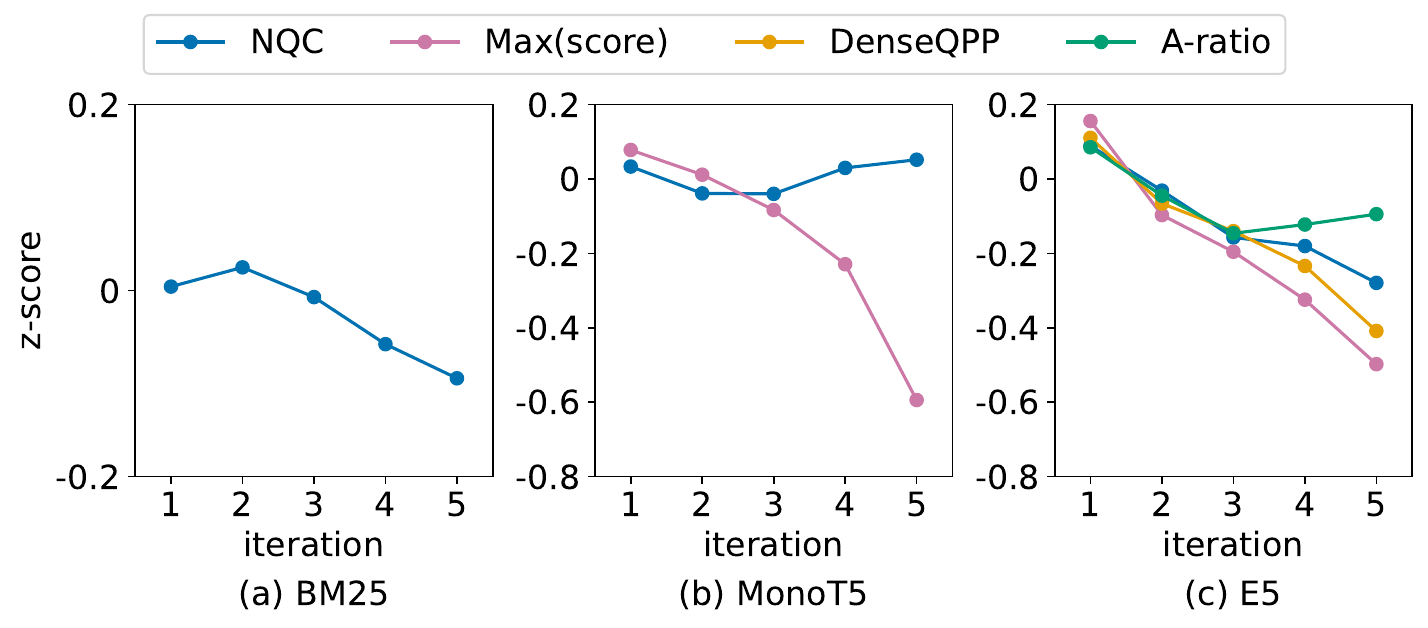}\vspace{-0.5em}
\caption{Average QPP estimates for the generated queries in each reasoning-retrieval iteration for Search-R1. QPP estimates are normalised by z-score.}\label{fig:search-r1-trend}\vspace{-1em}
\end{figure}

\begin{figure}[tb]
\includegraphics[width=85mm]{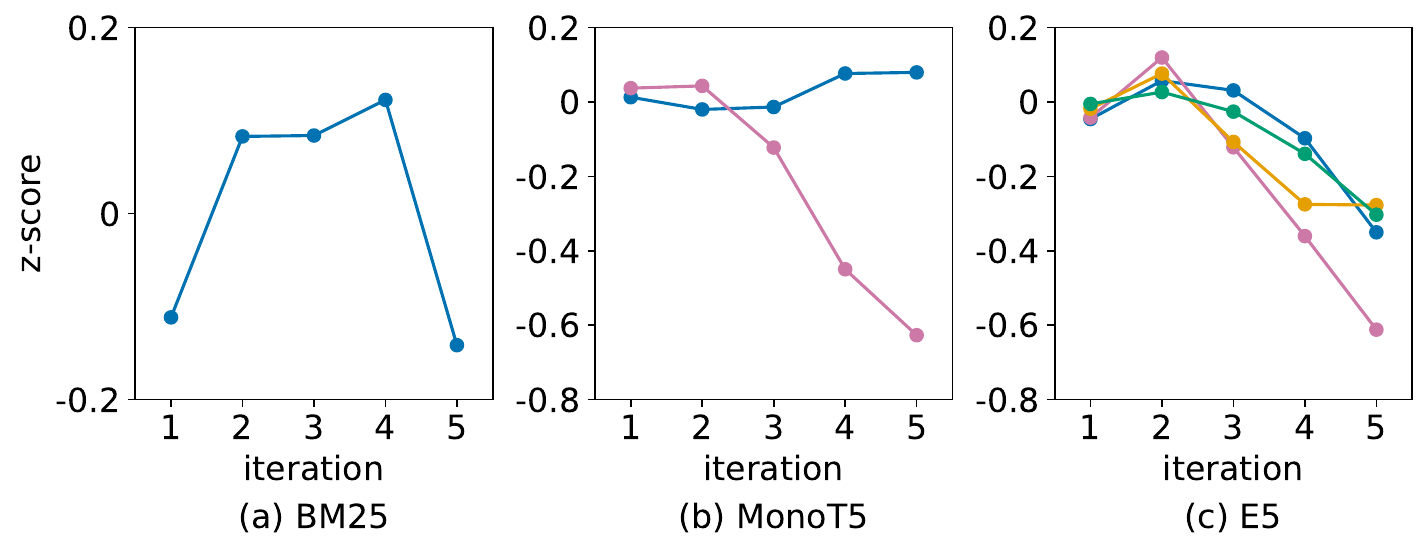}\vspace{-0.5em}
\caption{Average QPP estimates of generated queries in each reasoning-retrieval iteration for R1-Searcher, legend as per Figure~\ref{fig:search-r1-trend}.}\label{fig:r1-searcher-trend}\vspace{-1em}
\end{figure}



\vspace{-1mm}
\subsection{RQ-1: Effect of different retriever configurations on answer quality and number of iterations}

Table~\ref{table:system_summarisation} presents the average answer quality (measured in terms of EM and F1), and also the number of retrieval iterations (\numI{}) obtained with Search-R1 and R1-Searcher models using three different retriever configurations. The results show that for both Search-R1 and R1-Searcher, more effective retrievers -- MonoT5 and E5 -- generally lead to fewer reasoning-retrieval iterations \newcontent{initiated by the LLM} before the final answer is generated. 

\looseness -1 Additionally, the answers generated using results from more effective retrievers tend to be of higher quality. For example, Searcher-R1 achieves an F1 score of 0.4185 with an average of 2.52 iterations when using BM25 as the retriever. Using the E5 retriever leads to an increase in F1 score, to 0.5687, and this, in fact, is obtained with a reduced average reasoning length of 2.0 iterations. These findings suggest that retrieval effectiveness influences both the length of the reasoning process and the quality of the generated answers.

Figure~\ref{fig:distribution} provides a more detailed view on number of iterations, across the 3610 NQ test queries for both Search-R1 (a) and R1-Searcher (b). \newcontent{The dark part of each bar shows correct answers; the light part shows wrong answers.} From the plots of Figure~\ref{fig:distribution}, we observe that, for both of the Agentic RAG models, using BM25 as the retriever exhibit longer tails, indicating an extended reasoning process compared to models that use E5 and MonoT5 as the retriever. 


This observation raises a further question of whether the number of iterations (\numI{}) is correlated with average answer quality in Agentic RAG. Table~\ref{table:iteration_length_correlation} reports the Spearman's $\rho$ correlation between them within each RAG configuration. From the left side of the table, we observe a moderate negative correlation between \numI{} and F1 score for Search-R1 (i.e. more iterations implies lower likelihood of a good answer). For R1-Searcher, the correlations are weaker but remain negative suggesting that shorter reasoning processes tend to lead to higher answer quality.

\looseness -1 Another observation for Search-R1 is that a more effective retriever shows a stronger negative correlation, e.g., the correlation of both E5 and MonoT5 are lower than that of BM25. A plausible explanation for this is that more effective retrieval models return documents that are likely to be more relevant to the generated queries at each iteration, allowing the system to potentially generate better follow-up queries, and reach the final answer with fewer iterations.

\textbf{To conclude for RQ-1}, for both the Search-R1 and R1-Searcher Agentic RAG models, adopting more effective retrievers can shorten the reasoning process and lead to final answers of higher quality. We now investigate whether QPP estimates of query performance can provide any indication of answer quality.

    
\subsection{RQ-2: Predicted Generated Query Quality versus Iterations}

\looseness -1 Figures~\ref{fig:search-r1-trend} and \ref{fig:r1-searcher-trend} show the average QPP values for queries generated in the first five iterations across the test set. Each figure contains three sub-plots, one for each retrieval model. Different QPP predictors are applied for each retriever, based on their various applicabilities. Note that we normalise each predictor by z-score, to allow different predictors (with different ranges) to be more easily shown in a single plot. Finally, as noted earlier when analysing Figure~\ref{fig:distribution}, few questions have more than 5 iterations, so we show QPP values for iterations 1-5.

Overall, across both  Figures~\ref{fig:search-r1-trend} and \ref{fig:r1-searcher-trend}, we observe a decreasing trend in QPP estimates as the reasoning process progresses through iterations. This suggests that the quality of retrieval, in terms of relevance to the generated queries, generally deteriorates over iterations.
This overall downwards trend is consistent across the QPP methods evaluated on both Search-R1 and R1-Searcher pipelines. The only exception appears when NQC is applied to MonoT5, as shown in Figure~\ref{fig:search-r1-trend}(b) and Figure~\ref{fig:r1-searcher-trend}(b).

However, this exception is explainable, as NQC is not an accurate predictor for query performance in MonoT5. The score distribution produced by MonoT5, as argued by \citet{WRIG}, does not align with the assumptions underlying the statistical QPP methods - such as NQC. As a result of this, the NQC predictor may not reflect accurately the relevance or the usefulness of the documents retrieved for the purpose of LLM reasoning.


Additionally, these decreasing trends in the QPP estimates are consistent for the better performing QPP models, e.g., Max(Score) for Mono-T5. 
A likely explanation for this is that accurate QPP estimates correlate with the usefulness of the intermediate retrieved documents in answering the generated query. For instance, Max(Score) for Mono-T5 can be interpreted as an upper bound on the topical relevance between the query and the retrieved documents. If this score is low, the retrieved content is unlikely to satisfy the information need of the query -- potentially prolonging the overall reasoning process.


\textbf{To conclude for RQ-2}, based on the predictions by the QPP methods employed, it can be observed that the average quality of queries generated by the Agentic RAG models tends to decrease as the number of reasoning iterations increases. The most likely reason for this is that the queries for which the answer generation converges in a small number of iterations are likely to be more central to the information need of the input question (and hence expected to be of higher quality as estimated by a QPP model). On the other hand, for questions that end up taking a larger number of iterations, the generated answers are not of high quality (recall \newcontent{the decreasing EM along iterations shown in Figure~\ref{fig:distribution}}, and the negative correlations between the number of iterations and F1-score from Table \ref{table:iteration_length_correlation}). It is likely that the lower answer quality associated with an increase in the number of iterations is due to more noise being added to the generated text for reasoning itself by the low-quality retrievals. To better illustrate this phenomenon, we study the outputs of two example queries of Search-R1 in Appendix~\ref{sec:appendix}, \newcontent{which shows that low-quality retrievals for off-topic intermediate queries lead to a wrong answer with more iterations.}.


\subsection{RQ-3: Correlation between Predicted Query Performance and Answer Quality}

\begin{table}[tb]
\centering
\caption{Spearman's $\rho$ correlation between QPP estimation (for the query generated in the first iteration) and answer quality (measured by F1), for both models with various retrievers. A `-' value in the table denotes that a predictor is not applicable for a given retriever (as discussed in Section~\ref{ss:qpp_method_settings}). All observed correlations are significant.
}\label{table:qpp_correlation}

\begin{tabular}{@{}l ccc ccc@{}}
\toprule
\small
& \multicolumn{6}{c}{Agentic RAG Models}\\
\cmidrule(r){2-7}
& \multicolumn{3}{c}{Search-R1} & \multicolumn{3}{c}{R1-Searcher} \\
\cmidrule(r){2-4} \cmidrule(l){5-7}  
QPP & BM25 & MonoT5 & E5 & BM25 & MonoT5 & E5 \\
\midrule 
 NQC & .1297 & .0524 & .2394 & .1205 & .0096 & .0515 \\
   Max(Score) & - & .2383 & .2369 & - & .1735 & .0919  \\
 Dense-QPP & - & - & .1871 & - & - & .0834 \\
 A-Pair-Ratio & - & - & .2497 & - & - & .0297 \\
\bottomrule
\end{tabular}
\vspace{-1mm}
\end{table}

\looseness -1 We now to turn to examine if the QPP values of the queries are correlated to the final answer quality. In particular, we postulate that with high number of queries being answered in a single iteration (see Section~5.1), the retrieval quality of the first query is most likely to be indicative of the answer quality. Hence, we focus on the correlation between the QPP score of the first generated query and the F1 score of the final output answer. Table~\ref{table:qpp_correlation} reports the observed correlations.
From the table, we note that all correlations are consistently positive, suggesting that a higher QPP estimate for the first generated query tends to be associated with higher-quality answers. However, these correlations are weak, and less than what has been achievable for standard retrieval tasks (which can be as high as Spearman's $\rho > 0.5$)~\cite{WRIG, aPairRatio}. This emphasises that QPPs are not entirely capturing final answer quality.

Comparing the two Agentic RAG models, the observed correlations are higher for Search-R1 than for R1-Searcher. Considering the answer quality of these two Agentic RAG frameworks in Table~\ref{table:system_summarisation}, QPP estimates may be better correlated with the quality of answers generated by more effective RAG systems, in that Search-R1 makes better use of the retrieved documents, and hence is more susceptible to poor retrieval, leading to higher correlations.

When BM25 is used as the retriever, both pipelines show a weak but positive correlation using the NQC predictor, around 0.12–0.13. When the retrieval results are re-ranked by MonoT5, the correlation for NQC decreases (as expected because NQC is not an effective QPP model for neural re-rankers~\cite{WRIG}). In contrast, the QPP method Max(Score) achieves higher correlations, around 0.2. When E5 is used for retrieval, all QPP predictors in the Search-R1 pipeline show stronger correlations, particularly the coherence-based predictor A-ratio~\cite{aPairRatio}, achieving the highest correlation at 0.25. 

These observations indicate that QPP can serve as a useful signal in predicting the answer quality in the setting of Agentic RAG. However, in this initial study, we focus on correlating the answer quality with the QPP about the first generated query; higher correlation may be observed with improved aggregation methods of the QPP estimates in the multi-iteration reasoning process~\cite{thesisMariaV}.

\textbf{To conclude for RQ-3}, we found that QPP estimates of the queries generated in the first iteration of both Agentic RAG models are positively correlated with the quality of the final answer. Although the correlations are weak, they indicate promise worth further investigation.


\vspace{-1mm}
\section{Conclusions}\label{s:conclusion}

\looseness -1 In this work, we analysed the retrieval component in Agentic RAG models, using QPP as a tool to estimate the retrieval quality for queries generated by the Search-R1 and R1-Searcher models. Our experiments show that, for both these models, more effective retrieval models (e.g., E5) tend to yield higher-quality answers with fewer reasoning-retrieval iterations. Moreover, by evaluating QPP estimates of the generated queries at each iteration, we observe that retrieval quality tends to decrease over the course of the reasoning process. A case study in Appendix~\ref{sec:appendix} \newcontent{suggests that the ambiguity of the original query may be a reason leading to this decrease in retrieval quality.} Finally, we observe a weak positive correlation between the QPP estimations of the first generated query and the final answer quality. While the correlations are weak, it, nonetheless, provides a potentially useful signal in predicting the quality of generated answers. Overall, our results emphasise the importance of the first retrieval iteration in final answer quality.

\looseness -1 In terms of limitations, this study focused on just two models: Search-R1 and R1-Searcher, while other Agentic RAG models, such as Search-O1~\cite{searchO1} may exhibit different behaviours. Furthermore, our experiments are conducted only on the NQ dataset, while Agentic RAG models may be better suited to multi-hop QA questions, such as HotpotQA~\cite{hotpotQA}. We plan to address these limitations in a future extension.

\looseness -1 This work opens the doors to using IR techniques, including QPP, to improve the searching behaviour of Agentic RAG models. For instance, QPP may be used to adaptively decide whether the retrieval results are useful enough to show to the reasoning model; alternatively, QPP estimates could serve as a reward signal during the training time to help the model learn to execute better queries. This goes beyond the reward signals used in the training of R1-Searcher, which only reward the model for making search requests and generating responses that are validly formatted~\cite{r1Searcher}. We believe that, in the era of RAG, applying QPP to enhance the LLM's ability to better utilise IR retrieval models is a promising research direction.


\bibliographystyle{ACM-Reference-Format}

\onecolumn

\begin{thebibliography}{49}


\ifx \showCODEN    \undefined \def \showCODEN     #1{\unskip}     \fi
\ifx \showDOI      \undefined \def \showDOI       #1{#1}\fi
\ifx \showISBNx    \undefined \def \showISBNx     #1{\unskip}     \fi
\ifx \showISBNxiii \undefined \def \showISBNxiii  #1{\unskip}     \fi
\ifx \showISSN     \undefined \def \showISSN      #1{\unskip}     \fi
\ifx \showLCCN     \undefined \def \showLCCN      #1{\unskip}     \fi
\ifx \shownote     \undefined \def \shownote      #1{#1}          \fi
\ifx \showarticletitle \undefined \def \showarticletitle #1{#1}   \fi
\ifx \showURL      \undefined \def \showURL       {\relax}        \fi
\providecommand\bibfield[2]{#2}
\providecommand\bibinfo[2]{#2}
\providecommand\natexlab[1]{#1}
\providecommand\showeprint[2][]{arXiv:#2}

\bibitem[\protect\citeauthoryear{Alaofi, Arabzadeh, Clarke, and Sanderson}{Alaofi et~al\mbox{.}}{2024}]%
        {evalGenIR}
\bibfield{author}{\bibinfo{person}{Marwah Alaofi}, \bibinfo{person}{Negar Arabzadeh}, \bibinfo{person}{Charles L.~A. Clarke}, {and} \bibinfo{person}{Mark Sanderson}.} \bibinfo{year}{2024}\natexlab{}.
\newblock \bibinfo{title}{{Generative Information Retrieval Evaluation}}.
\newblock
\newblock
\showeprint[arxiv]{2404.08137}~[cs.IR]
\urldef\tempurl%
\url{https://arxiv.org/abs/2404.08137}
\showURL{%
\tempurl}


\bibitem[\protect\citeauthoryear{Arabzadeh, Bigdeli, Zihayat, and Bagheri}{Arabzadeh et~al\mbox{.}}{2021a}]%
        {coherenceQPP}
\bibfield{author}{\bibinfo{person}{Negar Arabzadeh}, \bibinfo{person}{Amin Bigdeli}, \bibinfo{person}{Morteza Zihayat}, {and} \bibinfo{person}{Ebrahim Bagheri}.} \bibinfo{year}{2021}\natexlab{a}.
\newblock \showarticletitle{Query Performance Prediction Through Retrieval Coherency}. In \bibinfo{booktitle}{\emph{Advances in Information Retrieval: 43rd European Conference on IR Research, ECIR 2021, Virtual Event, March 28 – April 1, 2021, Proceedings, Part II}}. \bibinfo{publisher}{Springer-Verlag}, \bibinfo{address}{Berlin, Heidelberg}, \bibinfo{pages}{193–200}.
\newblock
\showISBNx{978-3-030-72239-5}
\urldef\tempurl%
\url{https://doi.org/10.1007/978-3-030-72240-1_15}
\showURL{%
\tempurl}


\bibitem[\protect\citeauthoryear{Arabzadeh, Khodabakhsh, and Bagheri}{Arabzadeh et~al\mbox{.}}{2021b}]%
        {bertQPP}
\bibfield{author}{\bibinfo{person}{Negar Arabzadeh}, \bibinfo{person}{Maryam Khodabakhsh}, {and} \bibinfo{person}{Ebrahim Bagheri}.} \bibinfo{year}{2021}\natexlab{b}.
\newblock \showarticletitle{BERT-QPP: Contextualized Pre-trained transformers for Query Performance Prediction}. In \bibinfo{booktitle}{\emph{Proceedings of the 30th ACM International Conference on Information \& Knowledge Management}} (Virtual Event, Queensland, Australia) \emph{(\bibinfo{series}{CIKM '21})}. \bibinfo{publisher}{Association for Computing Machinery}, \bibinfo{address}{New York, NY, USA}, \bibinfo{pages}{2857–2861}.
\newblock
\showISBNx{9781450384469}
\urldef\tempurl%
\url{https://doi.org/10.1145/3459637.3482063}
\showURL{%
\tempurl}


\bibitem[\protect\citeauthoryear{Brown, Mann, Ryder, Subbiah, Kaplan, Dhariwal, Neelakantan, Shyam, Sastry, Askell, Agarwal, Herbert-Voss, Krueger, Henighan, Child, Ramesh, Ziegler, Wu, Winter, Hesse, Chen, Sigler, Litwin, Gray, Chess, Clark, Berner, McCandlish, Radford, Sutskever, and Amodei}{Brown et~al\mbox{.}}{2020}]%
        {fewShotLearnerLLM}
\bibfield{author}{\bibinfo{person}{Tom~B. Brown}, \bibinfo{person}{Benjamin Mann}, \bibinfo{person}{Nick Ryder}, \bibinfo{person}{Melanie Subbiah}, \bibinfo{person}{Jared Kaplan}, \bibinfo{person}{Prafulla Dhariwal}, \bibinfo{person}{Arvind Neelakantan}, \bibinfo{person}{Pranav Shyam}, \bibinfo{person}{Girish Sastry}, \bibinfo{person}{Amanda Askell}, \bibinfo{person}{Sandhini Agarwal}, \bibinfo{person}{Ariel Herbert-Voss}, \bibinfo{person}{Gretchen Krueger}, \bibinfo{person}{Tom Henighan}, \bibinfo{person}{Rewon Child}, \bibinfo{person}{Aditya Ramesh}, \bibinfo{person}{Daniel~M. Ziegler}, \bibinfo{person}{Jeffrey Wu}, \bibinfo{person}{Clemens Winter}, \bibinfo{person}{Christopher Hesse}, \bibinfo{person}{Mark Chen}, \bibinfo{person}{Eric Sigler}, \bibinfo{person}{Mateusz Litwin}, \bibinfo{person}{Scott Gray}, \bibinfo{person}{Benjamin Chess}, \bibinfo{person}{Jack Clark}, \bibinfo{person}{Christopher Berner}, \bibinfo{person}{Sam McCandlish}, \bibinfo{person}{Alec Radford}, \bibinfo{person}{Ilya Sutskever},
  {and} \bibinfo{person}{Dario Amodei}.} \bibinfo{year}{2020}\natexlab{}.
\newblock \showarticletitle{Language models are few-shot learners}. In \bibinfo{booktitle}{\emph{Proceedings of the 34th International Conference on Neural Information Processing Systems}} \emph{(\bibinfo{series}{NIPS '20})}. \bibinfo{publisher}{Curran Associates Inc.}, \bibinfo{address}{Red Hook, NY, USA}, Article \bibinfo{articleno}{159}.
\newblock
\showISBNx{9781713829546}


\bibitem[\protect\citeauthoryear{Chen, Stanovsky, Singh, and Gardner}{Chen et~al\mbox{.}}{2019}]%
        {evaluatingQAEval}
\bibfield{author}{\bibinfo{person}{Anthony Chen}, \bibinfo{person}{Gabriel Stanovsky}, \bibinfo{person}{Sameer Singh}, {and} \bibinfo{person}{Matt Gardner}.} \bibinfo{year}{2019}\natexlab{}.
\newblock \showarticletitle{Evaluating Question Answering Evaluation}. In \bibinfo{booktitle}{\emph{Proceedings of the 2nd Workshop on Machine Reading for Question Answering}}. \bibinfo{publisher}{Association for Computational Linguistics}, \bibinfo{address}{Hong Kong, China}, \bibinfo{pages}{119--124}.
\newblock
\urldef\tempurl%
\url{https://aclanthology.org/D19-5817/}
\showURL{%
\tempurl}


\bibitem[\protect\citeauthoryear{Cronen-Townsend, Zhou, and Croft}{Cronen-Townsend et~al\mbox{.}}{2002}]%
        {Clarity}
\bibfield{author}{\bibinfo{person}{Steve Cronen-Townsend}, \bibinfo{person}{Yun Zhou}, {and} \bibinfo{person}{W.~Bruce Croft}.} \bibinfo{year}{2002}\natexlab{}.
\newblock \showarticletitle{Predicting query performance} \emph{(\bibinfo{series}{SIGIR '02})}. \bibinfo{publisher}{Association for Computing Machinery}, \bibinfo{address}{New York, NY, USA}, \bibinfo{pages}{299–306}.
\newblock
\showISBNx{1581135610}
\urldef\tempurl%
\url{https://doi.org/10.1145/564376.564429}
\showURL{%
\tempurl}


\bibitem[\protect\citeauthoryear{Cuconasu, Trappolini, Siciliano, Filice, Campagnano, Maarek, Tonellotto, and Silvestri}{Cuconasu et~al\mbox{.}}{2024}]%
        {powerOfNoise}
\bibfield{author}{\bibinfo{person}{Florin Cuconasu}, \bibinfo{person}{Giovanni Trappolini}, \bibinfo{person}{Federico Siciliano}, \bibinfo{person}{Simone Filice}, \bibinfo{person}{Cesare Campagnano}, \bibinfo{person}{Yoelle Maarek}, \bibinfo{person}{Nicola Tonellotto}, {and} \bibinfo{person}{Fabrizio Silvestri}.} \bibinfo{year}{2024}\natexlab{}.
\newblock \showarticletitle{{The Power of Noise: Redefining Retrieval for RAG Systems}}. In \bibinfo{booktitle}{\emph{Proceedings of the 47th International ACM SIGIR Conference on Research and Development in Information Retrieval}} (Washington DC, USA) \emph{(\bibinfo{series}{SIGIR '24})}. \bibinfo{publisher}{Association for Computing Machinery}, \bibinfo{address}{New York, NY, USA}, \bibinfo{pages}{719–729}.
\newblock
\showISBNx{9798400704314}
\urldef\tempurl%
\url{https://doi.org/10.1145/3626772.3657834}
\showURL{%
\tempurl}


\bibitem[\protect\citeauthoryear{Datta, Ganguly, Mitra, and Greene}{Datta et~al\mbox{.}}{2022a}]%
        {WRIG}
\bibfield{author}{\bibinfo{person}{Suchana Datta}, \bibinfo{person}{Debasis Ganguly}, \bibinfo{person}{Mandar Mitra}, {and} \bibinfo{person}{Derek Greene}.} \bibinfo{year}{2022}\natexlab{a}.
\newblock \showarticletitle{A Relative Information Gain-based Query Performance Prediction Framework with Generated Query Variants}.
\newblock \bibinfo{journal}{\emph{ACM Trans. Inf. Syst.}} \bibinfo{volume}{41}, \bibinfo{number}{2}, Article \bibinfo{articleno}{38} (\bibinfo{date}{dec} \bibinfo{year}{2022}), \bibinfo{numpages}{31}~pages.
\newblock
\showISSN{1046-8188}
\urldef\tempurl%
\url{https://doi.org/10.1145/3545112}
\showURL{%
\tempurl}


\bibitem[\protect\citeauthoryear{Datta, MacAvaney, Ganguly, and Greene}{Datta et~al\mbox{.}}{2022b}]%
        {qppBERT-PL}
\bibfield{author}{\bibinfo{person}{Suchana Datta}, \bibinfo{person}{Sean MacAvaney}, \bibinfo{person}{Debasis Ganguly}, {and} \bibinfo{person}{Derek Greene}.} \bibinfo{year}{2022}\natexlab{b}.
\newblock \showarticletitle{A 'Pointwise-Query, Listwise-Document' based Query Performance Prediction Approach}. In \bibinfo{booktitle}{\emph{Proceedings of the 45th International ACM SIGIR Conference on Research and Development in Information Retrieval}} (Madrid, Spain) \emph{(\bibinfo{series}{SIGIR '22})}. \bibinfo{publisher}{Association for Computing Machinery}, \bibinfo{address}{New York, NY, USA}, \bibinfo{pages}{2148–2153}.
\newblock
\showISBNx{9781450387323}
\urldef\tempurl%
\url{https://doi.org/10.1145/3477495.3531821}
\showURL{%
\tempurl}


\bibitem[\protect\citeauthoryear{Dubey et~al\mbox{.}}{Dubey et~al\mbox{.}}{2024}]%
        {llama3}
\bibfield{author}{\bibinfo{person}{Abhimanyu Dubey} {et~al\mbox{.}}} \bibinfo{year}{2024}\natexlab{}.
\newblock \bibinfo{title}{The {Llama} 3 Herd of Models}.
\newblock
\newblock
\showeprint[arxiv]{2407.21783}~[cs.AI]
\urldef\tempurl%
\url{https://arxiv.org/abs/2407.21783}
\showURL{%
\tempurl}


\bibitem[\protect\citeauthoryear{Dziri, Milton, Yu, Zaiane, and Reddy}{Dziri et~al\mbox{.}}{2022}]%
        {originOfHallucinations}
\bibfield{author}{\bibinfo{person}{Nouha Dziri}, \bibinfo{person}{Sivan Milton}, \bibinfo{person}{Mo Yu}, \bibinfo{person}{Osmar Zaiane}, {and} \bibinfo{person}{Siva Reddy}.} \bibinfo{year}{2022}\natexlab{}.
\newblock \showarticletitle{On the Origin of Hallucinations in Conversational Models: Is it the Datasets or the Models?}. In \bibinfo{booktitle}{\emph{Proceedings of the 2022 Conference of the North American Chapter of the Association for Computational Linguistics: Human Language Technologies}}. \bibinfo{publisher}{Association for Computational Linguistics}, \bibinfo{address}{Seattle, United States}, \bibinfo{pages}{5271--5285}.
\newblock
\urldef\tempurl%
\url{https://aclanthology.org/2022.naacl-main.387/}
\showURL{%
\tempurl}


\bibitem[\protect\citeauthoryear{Faggioli, Ferro, Muntean, Perego, and Tonellotto}{Faggioli et~al\mbox{.}}{2023}]%
        {denseQPP}
\bibfield{author}{\bibinfo{person}{Guglielmo Faggioli}, \bibinfo{person}{Nicola Ferro}, \bibinfo{person}{Cristina~Ioana Muntean}, \bibinfo{person}{Raffaele Perego}, {and} \bibinfo{person}{Nicola Tonellotto}.} \bibinfo{year}{2023}\natexlab{}.
\newblock \showarticletitle{A Geometric Framework for Query Performance Prediction in Conversational Search}. In \bibinfo{booktitle}{\emph{Proceedings of the 46th International ACM SIGIR Conference on Research and Development in Information Retrieval}} (Taipei, Taiwan) \emph{(\bibinfo{series}{SIGIR '23})}. \bibinfo{publisher}{Association for Computing Machinery}, \bibinfo{address}{New York, NY, USA}, \bibinfo{pages}{1355–1365}.
\newblock
\showISBNx{9781450394086}
\urldef\tempurl%
\url{https://doi.org/10.1145/3539618.3591625}
\showURL{%
\tempurl}


\bibitem[\protect\citeauthoryear{Fang, Meng, and Macdonald}{Fang et~al\mbox{.}}{2025}]%
        {fang2025kirag}
\bibfield{author}{\bibinfo{person}{Jinyuan Fang}, \bibinfo{person}{Zaiqiao Meng}, {and} \bibinfo{person}{Craig Macdonald}.} \bibinfo{year}{2025}\natexlab{}.
\newblock \showarticletitle{KiRAG: Knowledge-Driven Iterative Retriever for Enhancing Retrieval-Augmented Generation}.
\newblock \bibinfo{journal}{\emph{arXiv preprint arXiv:2502.18397}} (\bibinfo{year}{2025}).
\newblock


\bibitem[\protect\citeauthoryear{Farquhar, Kossen, Kuhn, and Gal}{Farquhar et~al\mbox{.}}{2024}]%
        {detectHallucinations}
\bibfield{author}{\bibinfo{person}{Sebastian Farquhar}, \bibinfo{person}{Jannik Kossen}, \bibinfo{person}{Lorenz Kuhn}, {and} \bibinfo{person}{Yarin Gal}.} \bibinfo{year}{2024}\natexlab{}.
\newblock \showarticletitle{Detecting hallucinations in large language models using semantic entropy}.
\newblock \bibinfo{journal}{\emph{Nature (London)}} \bibinfo{volume}{630}, \bibinfo{number}{8017} (\bibinfo{year}{2024}), \bibinfo{pages}{625--630}.
\newblock
\showISBNx{0028-0836;1476-4687;}


\bibitem[\protect\citeauthoryear{Gao, Xiong, Gao, Jia, Pan, Bi, Dai, Sun, Wang, and Wang}{Gao et~al\mbox{.}}{2024}]%
        {ragReview}
\bibfield{author}{\bibinfo{person}{Yunfan Gao}, \bibinfo{person}{Yun Xiong}, \bibinfo{person}{Xinyu Gao}, \bibinfo{person}{Kangxiang Jia}, \bibinfo{person}{Jinliu Pan}, \bibinfo{person}{Yuxi Bi}, \bibinfo{person}{Yi Dai}, \bibinfo{person}{Jiawei Sun}, \bibinfo{person}{Meng Wang}, {and} \bibinfo{person}{Haofen Wang}.} \bibinfo{year}{2024}\natexlab{}.
\newblock \bibinfo{title}{Retrieval-Augmented Generation for Large Language Models: A Survey}.
\newblock
\newblock
\showeprint[arxiv]{2312.10997}~[cs.CL]
\urldef\tempurl%
\url{https://arxiv.org/abs/2312.10997}
\showURL{%
\tempurl}


\bibitem[\protect\citeauthoryear{Hauff, Hiemstra, and de~Jong}{Hauff et~al\mbox{.}}{2008}]%
        {preRetrievalSurvey}
\bibfield{author}{\bibinfo{person}{Claudia Hauff}, \bibinfo{person}{Djoerd Hiemstra}, {and} \bibinfo{person}{Franciska de Jong}.} \bibinfo{year}{2008}\natexlab{}.
\newblock \showarticletitle{A survey of pre-retrieval query performance predictors}. In \bibinfo{booktitle}{\emph{Proceedings of the 17th ACM Conference on Information and Knowledge Management}} (Napa Valley, California, USA) \emph{(\bibinfo{series}{CIKM '08})}. \bibinfo{pages}{1419–1420}.
\newblock
\urldef\tempurl%
\url{https://doi.org/10.1145/1458082.1458311}
\showURL{%
\tempurl}


\bibitem[\protect\citeauthoryear{He, Chen, He, Yan, Wei, Luo, and Ling}{He et~al\mbox{.}}{2024}]%
        {covRAG}
\bibfield{author}{\bibinfo{person}{Bolei He}, \bibinfo{person}{Nuo Chen}, \bibinfo{person}{Xinran He}, \bibinfo{person}{Lingyong Yan}, \bibinfo{person}{Zhenkai Wei}, \bibinfo{person}{Jinchang Luo}, {and} \bibinfo{person}{Zhen-Hua Ling}.} \bibinfo{year}{2024}\natexlab{}.
\newblock \showarticletitle{Retrieving, Rethinking and Revising: The Chain-of-Verification Can Improve Retrieval Augmented Generation}. In \bibinfo{booktitle}{\emph{Findings of the Association for Computational Linguistics: EMNLP 2024}}. \bibinfo{publisher}{Association for Computational Linguistics}, \bibinfo{address}{Miami, Florida, USA}, \bibinfo{pages}{10371--10393}.
\newblock
\urldef\tempurl%
\url{https://aclanthology.org/2024.findings-emnlp.607/}
\showURL{%
\tempurl}


\bibitem[\protect\citeauthoryear{He and Ounis}{He and Ounis}{2004}]%
        {spire2004}
\bibfield{author}{\bibinfo{person}{Ben He} {and} \bibinfo{person}{Iadh Ounis}.} \bibinfo{year}{2004}\natexlab{}.
\newblock \showarticletitle{Inferring Query Performance Using Pre-retrieval Predictors}. In \bibinfo{booktitle}{\emph{String Processing and Information Retrieval}}. \bibinfo{pages}{43--54}.
\newblock
\showISBNx{978-3-540-30213-1}


\bibitem[\protect\citeauthoryear{Ho, Duong~Nguyen, Sugawara, and Aizawa}{Ho et~al\mbox{.}}{2020}]%
        {2WikiMultihopQA}
\bibfield{author}{\bibinfo{person}{Xanh Ho}, \bibinfo{person}{Anh-Khoa Duong~Nguyen}, \bibinfo{person}{Saku Sugawara}, {and} \bibinfo{person}{Akiko Aizawa}.} \bibinfo{year}{2020}\natexlab{}.
\newblock \showarticletitle{Constructing A Multi-hop {QA} Dataset for Comprehensive Evaluation of Reasoning Steps}. In \bibinfo{booktitle}{\emph{Proceedings of the 28th International Conference on Computational Linguistics}}. \bibinfo{pages}{6609--6625}.
\newblock
\urldef\tempurl%
\url{https://aclanthology.org/2020.coling-main.580/}
\showURL{%
\tempurl}


\bibitem[\protect\citeauthoryear{Izacard and Grave}{Izacard and Grave}{2021}]%
        {FiD}
\bibfield{author}{\bibinfo{person}{Gautier Izacard} {and} \bibinfo{person}{Edouard Grave}.} \bibinfo{year}{2021}\natexlab{}.
\newblock \showarticletitle{Leveraging Passage Retrieval with Generative Models for Open Domain Question Answering}. In \bibinfo{booktitle}{\emph{Proceedings of the 16th Conference of the European Chapter of the Association for Computational Linguistics: Main Volume}}. \bibinfo{publisher}{Association for Computational Linguistics}, \bibinfo{address}{Online}, \bibinfo{pages}{874--880}.
\newblock
\urldef\tempurl%
\url{https://aclanthology.org/2021.eacl-main.74/}
\showURL{%
\tempurl}


\bibitem[\protect\citeauthoryear{Izacard, Lewis, Lomeli, Hosseini, Petroni, Schick, Dwivedi-Yu, Joulin, Riedel, and Grave}{Izacard et~al\mbox{.}}{2023}]%
        {ATLAS}
\bibfield{author}{\bibinfo{person}{Gautier Izacard}, \bibinfo{person}{Patrick Lewis}, \bibinfo{person}{Maria Lomeli}, \bibinfo{person}{Lucas Hosseini}, \bibinfo{person}{Fabio Petroni}, \bibinfo{person}{Timo Schick}, \bibinfo{person}{Jane Dwivedi-Yu}, \bibinfo{person}{Armand Joulin}, \bibinfo{person}{Sebastian Riedel}, {and} \bibinfo{person}{Edouard Grave}.} \bibinfo{year}{2023}\natexlab{}.
\newblock \showarticletitle{Atlas: few-shot learning with retrieval augmented language models}.
\newblock \bibinfo{journal}{\emph{J. Mach. Learn. Res.}} \bibinfo{volume}{24}, \bibinfo{number}{1}, Article \bibinfo{articleno}{251} (\bibinfo{date}{Jan.} \bibinfo{year}{2023}), \bibinfo{numpages}{43}~pages.
\newblock
\showISSN{1532-4435}


\bibitem[\protect\citeauthoryear{Jiang, Xu, Gao, Sun, Liu, Dwivedi-Yu, Yang, Callan, and Neubig}{Jiang et~al\mbox{.}}{2023}]%
        {FLARE}
\bibfield{author}{\bibinfo{person}{Zhengbao Jiang}, \bibinfo{person}{Frank Xu}, \bibinfo{person}{Luyu Gao}, \bibinfo{person}{Zhiqing Sun}, \bibinfo{person}{Qian Liu}, \bibinfo{person}{Jane Dwivedi-Yu}, \bibinfo{person}{Yiming Yang}, \bibinfo{person}{Jamie Callan}, {and} \bibinfo{person}{Graham Neubig}.} \bibinfo{year}{2023}\natexlab{}.
\newblock \showarticletitle{Active Retrieval Augmented Generation}. In \bibinfo{booktitle}{\emph{Proceedings of the 2023 Conference on Empirical Methods in Natural Language Processing}}. \bibinfo{publisher}{Association for Computational Linguistics}, \bibinfo{address}{Singapore}, \bibinfo{pages}{7969--7992}.
\newblock
\urldef\tempurl%
\url{https://doi.org/10.18653/v1/2023.emnlp-main.495}
\showDOI{\tempurl}


\bibitem[\protect\citeauthoryear{Jin, Zeng, Yue, Yoon, Arik, Wang, Zamani, and Han}{Jin et~al\mbox{.}}{2025}]%
        {searchR1}
\bibfield{author}{\bibinfo{person}{Bowen Jin}, \bibinfo{person}{Hansi Zeng}, \bibinfo{person}{Zhenrui Yue}, \bibinfo{person}{Jinsung Yoon}, \bibinfo{person}{Sercan Arik}, \bibinfo{person}{Dong Wang}, \bibinfo{person}{Hamed Zamani}, {and} \bibinfo{person}{Jiawei Han}.} \bibinfo{year}{2025}\natexlab{}.
\newblock \bibinfo{title}{Search-R1: Training LLMs to Reason and Leverage Search Engines with Reinforcement Learning}.
\newblock
\newblock
\showeprint[arxiv]{2503.09516}~[cs.CL]
\urldef\tempurl%
\url{https://arxiv.org/abs/2503.09516}
\showURL{%
\tempurl}


\bibitem[\protect\citeauthoryear{Joshi, Sarwar, Varshney, Nag, Agrawal, and Naik}{Joshi et~al\mbox{.}}{2024}]%
        {reasoningPlaningRAG}
\bibfield{author}{\bibinfo{person}{Ashutosh Joshi}, \bibinfo{person}{Sheikh~Muhammad Sarwar}, \bibinfo{person}{Samarth Varshney}, \bibinfo{person}{Sreyashi Nag}, \bibinfo{person}{Shrivats Agrawal}, {and} \bibinfo{person}{Juhi Naik}.} \bibinfo{year}{2024}\natexlab{}.
\newblock \showarticletitle{REAPER: Reasoning based Retrieval Planning for Complex RAG Systems}. In \bibinfo{booktitle}{\emph{Proceedings of the 33rd ACM International Conference on Information and Knowledge Management}} (Boise, ID, USA) \emph{(\bibinfo{series}{CIKM '24})}. \bibinfo{publisher}{Association for Computing Machinery}, \bibinfo{address}{New York, NY, USA}, \bibinfo{pages}{4621–4628}.
\newblock
\showISBNx{9798400704369}
\urldef\tempurl%
\url{https://doi.org/10.1145/3627673.3680087}
\showURL{%
\tempurl}


\bibitem[\protect\citeauthoryear{Kwiatkowski, Palomaki, Redfield, Collins, Parikh, Alberti, Epstein, Polosukhin, Devlin, Lee, Toutanova, Jones, Kelcey, Chang, Dai, Uszkoreit, Le, and Petrov}{Kwiatkowski et~al\mbox{.}}{2019}]%
        {NQ}
\bibfield{author}{\bibinfo{person}{Tom Kwiatkowski}, \bibinfo{person}{Jennimaria Palomaki}, \bibinfo{person}{Olivia Redfield}, \bibinfo{person}{Michael Collins}, \bibinfo{person}{Ankur Parikh}, \bibinfo{person}{Chris Alberti}, \bibinfo{person}{Danielle Epstein}, \bibinfo{person}{Illia Polosukhin}, \bibinfo{person}{Jacob Devlin}, \bibinfo{person}{Kenton Lee}, \bibinfo{person}{Kristina Toutanova}, \bibinfo{person}{Llion Jones}, \bibinfo{person}{Matthew Kelcey}, \bibinfo{person}{Ming-Wei Chang}, \bibinfo{person}{Andrew~M. Dai}, \bibinfo{person}{Jakob Uszkoreit}, \bibinfo{person}{Quoc Le}, {and} \bibinfo{person}{Slav Petrov}.} \bibinfo{year}{2019}\natexlab{}.
\newblock \showarticletitle{Natural Questions: A Benchmark for Question Answering Research}.
\newblock \bibinfo{journal}{\emph{Transactions of the Association for Computational Linguistics}}  \bibinfo{volume}{7} (\bibinfo{year}{2019}), \bibinfo{pages}{452--466}.
\newblock
\urldef\tempurl%
\url{https://aclanthology.org/Q19-1026/}
\showURL{%
\tempurl}


\bibitem[\protect\citeauthoryear{Lewis, Perez, Piktus, Petroni, Karpukhin, Goyal, K\"{u}ttler, Lewis, Yih, Rockt\"{a}schel, Riedel, and Kiela}{Lewis et~al\mbox{.}}{2020}]%
        {ragInKnowledgeIntensiveNLP}
\bibfield{author}{\bibinfo{person}{Patrick Lewis}, \bibinfo{person}{Ethan Perez}, \bibinfo{person}{Aleksandra Piktus}, \bibinfo{person}{Fabio Petroni}, \bibinfo{person}{Vladimir Karpukhin}, \bibinfo{person}{Naman Goyal}, \bibinfo{person}{Heinrich K\"{u}ttler}, \bibinfo{person}{Mike Lewis}, \bibinfo{person}{Wen-tau Yih}, \bibinfo{person}{Tim Rockt\"{a}schel}, \bibinfo{person}{Sebastian Riedel}, {and} \bibinfo{person}{Douwe Kiela}.} \bibinfo{year}{2020}\natexlab{}.
\newblock \showarticletitle{Retrieval-augmented generation for knowledge-intensive {NLP} tasks}. In \bibinfo{booktitle}{\emph{Proceedings of the 34th International Conference on Neural Information Processing Systems}} (Vancouver, BC, Canada) \emph{(\bibinfo{series}{NIPS '20})}. \bibinfo{publisher}{Curran Associates Inc.}, \bibinfo{address}{Red Hook, NY, USA}, Article \bibinfo{articleno}{793}, \bibinfo{numpages}{16}~pages.
\newblock
\showISBNx{9781713829546}
\urldef\tempurl%
\url{https://arxiv.org/abs/2005.11401}
\showURL{%
\tempurl}


\bibitem[\protect\citeauthoryear{Li, Dong, Jin, Zhang, Zhou, Zhu, Zhang, and Dou}{Li et~al\mbox{.}}{2025}]%
        {searchO1}
\bibfield{author}{\bibinfo{person}{Xiaoxi Li}, \bibinfo{person}{Guanting Dong}, \bibinfo{person}{Jiajie Jin}, \bibinfo{person}{Yuyao Zhang}, \bibinfo{person}{Yujia Zhou}, \bibinfo{person}{Yutao Zhu}, \bibinfo{person}{Peitian Zhang}, {and} \bibinfo{person}{Zhicheng Dou}.} \bibinfo{year}{2025}\natexlab{}.
\newblock \bibinfo{title}{Search-o1: Agentic Search-Enhanced Large Reasoning Models}.
\newblock
\newblock
\showeprint[arxiv]{2501.05366}~[cs.AI]
\urldef\tempurl%
\url{https://arxiv.org/abs/2501.05366}
\showURL{%
\tempurl}


\bibitem[\protect\citeauthoryear{Liu, Lin, Hewitt, Paranjape, Bevilacqua, Petroni, and Liang}{Liu et~al\mbox{.}}{2024}]%
        {lostInTheMiddle}
\bibfield{author}{\bibinfo{person}{Nelson~F. Liu}, \bibinfo{person}{Kevin Lin}, \bibinfo{person}{John Hewitt}, \bibinfo{person}{Ashwin Paranjape}, \bibinfo{person}{Michele Bevilacqua}, \bibinfo{person}{Fabio Petroni}, {and} \bibinfo{person}{Percy Liang}.} \bibinfo{year}{2024}\natexlab{}.
\newblock \showarticletitle{Lost in the Middle: How Language Models Use Long Contexts}.
\newblock \bibinfo{journal}{\emph{Transactions of the Association for Computational Linguistics}}  \bibinfo{volume}{12} (\bibinfo{year}{2024}), \bibinfo{pages}{157--173}.
\newblock
\urldef\tempurl%
\url{https://doi.org/10.1162/tacl_a_00638}
\showDOI{\tempurl}


\bibitem[\protect\citeauthoryear{Macdonald, Fang, Parry, and Meng}{Macdonald et~al\mbox{.}}{2025}]%
        {macdonald25pyterrier_rag}
\bibfield{author}{\bibinfo{person}{Craig Macdonald}, \bibinfo{person}{Jinyuan Fang}, \bibinfo{person}{Andrew Parry}, {and} \bibinfo{person}{Zaiqiao Meng}.} \bibinfo{year}{2025}\natexlab{}.
\newblock \showarticletitle{{Constructing and Evaluating Declarative RAG Pipelines in PyTerrier}}. In \bibinfo{booktitle}{\emph{Proceedings of SIGIR 2025}}.
\newblock


\bibitem[\protect\citeauthoryear{Macdonald, Tonellotto, MacAvaney, and Ounis}{Macdonald et~al\mbox{.}}{2021}]%
        {pyterrier}
\bibfield{author}{\bibinfo{person}{Craig Macdonald}, \bibinfo{person}{Nicola Tonellotto}, \bibinfo{person}{Sean MacAvaney}, {and} \bibinfo{person}{Iadh Ounis}.} \bibinfo{year}{2021}\natexlab{}.
\newblock \showarticletitle{{PyTerrier}: Declarative Experimentation in Python from BM25 to Dense Retrieval}. In \bibinfo{booktitle}{\emph{Proceedings of the 30th ACM International Conference on Information \& Knowledge Management}} (Virtual Event, Queensland, Australia) \emph{(\bibinfo{series}{CIKM '21})}. \bibinfo{pages}{4526–4533}.
\newblock
\urldef\tempurl%
\url{https://doi.org/10.1145/3459637.3482013}
\showDOI{\tempurl}


\bibitem[\protect\citeauthoryear{Meng, Arabzadeh, Aliannejadi, and de~Rijke}{Meng et~al\mbox{.}}{2023}]%
        {qppFromAdHocToConvSearch}
\bibfield{author}{\bibinfo{person}{Chuan Meng}, \bibinfo{person}{Negar Arabzadeh}, \bibinfo{person}{Mohammad Aliannejadi}, {and} \bibinfo{person}{Maarten de Rijke}.} \bibinfo{year}{2023}\natexlab{}.
\newblock \showarticletitle{Query Performance Prediction: From Ad-hoc to Conversational Search}. In \bibinfo{booktitle}{\emph{Proceedings of the 46th International ACM SIGIR Conference on Research and Development in Information Retrieval}} (Taipei, Taiwan) \emph{(\bibinfo{series}{SIGIR '23})}. \bibinfo{publisher}{Association for Computing Machinery}, \bibinfo{address}{New York, NY, USA}, \bibinfo{pages}{2583–2593}.
\newblock
\showISBNx{9781450394086}
\urldef\tempurl%
\url{https://doi.org/10.1145/3539618.3591919}
\showURL{%
\tempurl}


\bibitem[\protect\citeauthoryear{Nogueira, Jiang, Pradeep, and Lin}{Nogueira et~al\mbox{.}}{2020}]%
        {monot5}
\bibfield{author}{\bibinfo{person}{Rodrigo~Frassetto Nogueira}, \bibinfo{person}{Zhiying Jiang}, \bibinfo{person}{Ronak Pradeep}, {and} \bibinfo{person}{Jimmy Lin}.} \bibinfo{year}{2020}\natexlab{}.
\newblock \showarticletitle{Document Ranking with a Pretrained Sequence-to-Sequence Model}. In \bibinfo{booktitle}{\emph{Findings of the Association for Computational Linguistics: {EMNLP} 2020, Online Event, 16-20 November 2020}} \emph{(\bibinfo{series}{Findings of {ACL}}, Vol.~\bibinfo{volume}{{EMNLP} 2020})}. \bibinfo{publisher}{Association for Computational Linguistics}, \bibinfo{pages}{708--718}.
\newblock


\bibitem[\protect\citeauthoryear{Parvez, Ahmad, Chakraborty, Ray, and Chang}{Parvez et~al\mbox{.}}{2021}]%
        {ragCodeGenAndSummarisation}
\bibfield{author}{\bibinfo{person}{Md~Rizwan Parvez}, \bibinfo{person}{Wasi Ahmad}, \bibinfo{person}{Saikat Chakraborty}, \bibinfo{person}{Baishakhi Ray}, {and} \bibinfo{person}{Kai-Wei Chang}.} \bibinfo{year}{2021}\natexlab{}.
\newblock \showarticletitle{Retrieval Augmented Code Generation and Summarization}. In \bibinfo{booktitle}{\emph{Findings of the Association for Computational Linguistics: EMNLP 2021}}. \bibinfo{publisher}{Association for Computational Linguistics}, \bibinfo{address}{Punta Cana, Dominican Republic}, \bibinfo{pages}{2719--2734}.
\newblock
\urldef\tempurl%
\url{https://aclanthology.org/2021.findings-emnlp.232/}
\showURL{%
\tempurl}


\bibitem[\protect\citeauthoryear{Roy, Ribeiro, Blloshmi, and Small}{Roy et~al\mbox{.}}{2024}]%
        {selfRAG}
\bibfield{author}{\bibinfo{person}{Nirmal Roy}, \bibinfo{person}{Leonardo F.~R. Ribeiro}, \bibinfo{person}{Rexhina Blloshmi}, {and} \bibinfo{person}{Kevin Small}.} \bibinfo{year}{2024}\natexlab{}.
\newblock \showarticletitle{Learning When to Retrieve, What to Rewrite, and How to Respond in Conversational {QA}}. In \bibinfo{booktitle}{\emph{Findings of the Association for Computational Linguistics: EMNLP 2024}}. \bibinfo{publisher}{Association for Computational Linguistics}, \bibinfo{address}{Miami, Florida, USA}, \bibinfo{pages}{10604--10625}.
\newblock
\urldef\tempurl%
\url{https://aclanthology.org/2024.findings-emnlp.622/}
\showURL{%
\tempurl}


\bibitem[\protect\citeauthoryear{Salemi and Zamani}{Salemi and Zamani}{2024}]%
        {ragRelevanceLabel}
\bibfield{author}{\bibinfo{person}{Alireza Salemi} {and} \bibinfo{person}{Hamed Zamani}.} \bibinfo{year}{2024}\natexlab{}.
\newblock \showarticletitle{Evaluating Retrieval Quality in Retrieval-Augmented Generation}. In \bibinfo{booktitle}{\emph{Proceedings of the 47th International ACM SIGIR Conference on Research and Development in Information Retrieval}} (Washington DC, USA) \emph{(\bibinfo{series}{SIGIR '24})}. \bibinfo{publisher}{Association for Computing Machinery}, \bibinfo{address}{New York, NY, USA}, \bibinfo{pages}{2395–2400}.
\newblock
\showISBNx{9798400704314}
\urldef\tempurl%
\url{https://doi.org/10.1145/3626772.3657957}
\showURL{%
\tempurl}


\bibitem[\protect\citeauthoryear{Shtok, Kurland, and Carmel}{Shtok et~al\mbox{.}}{2010}]%
        {UEF}
\bibfield{author}{\bibinfo{person}{Anna Shtok}, \bibinfo{person}{Oren Kurland}, {and} \bibinfo{person}{David Carmel}.} \bibinfo{year}{2010}\natexlab{}.
\newblock \showarticletitle{Using statistical decision theory and relevance models for query-performance prediction}. In \bibinfo{booktitle}{\emph{Proceedings of the 33rd International ACM SIGIR Conference on Research and Development in Information Retrieval}} (Geneva, Switzerland) \emph{(\bibinfo{series}{SIGIR '10})}. \bibinfo{pages}{259–266}.
\newblock
\urldef\tempurl%
\url{https://doi.org/10.1145/1835449.1835494}
\showDOI{\tempurl}


\bibitem[\protect\citeauthoryear{Shtok, Kurland, Carmel, Raiber, and Markovits}{Shtok et~al\mbox{.}}{2012}]%
        {NQC}
\bibfield{author}{\bibinfo{person}{Anna Shtok}, \bibinfo{person}{Oren Kurland}, \bibinfo{person}{David Carmel}, \bibinfo{person}{Fiana Raiber}, {and} \bibinfo{person}{Gad Markovits}.} \bibinfo{year}{2012}\natexlab{}.
\newblock \showarticletitle{Predicting Query Performance by Query-Drift Estimation}.
\newblock \bibinfo{journal}{\emph{ACM Trans. Inf. Syst.}} \bibinfo{volume}{30}, \bibinfo{number}{2}, Article \bibinfo{articleno}{11} (\bibinfo{date}{May} \bibinfo{year}{2012}), \bibinfo{numpages}{35}~pages.
\newblock
\showISSN{1046-8188}
\urldef\tempurl%
\url{https://doi.org/10.1145/2180868.2180873}
\showDOI{\tempurl}


\bibitem[\protect\citeauthoryear{Singh, Ehtesham, Kumar, and Khoei}{Singh et~al\mbox{.}}{2025}]%
        {agenticRAGSurvey}
\bibfield{author}{\bibinfo{person}{Aditi Singh}, \bibinfo{person}{Abul Ehtesham}, \bibinfo{person}{Saket Kumar}, {and} \bibinfo{person}{Tala~Talaei Khoei}.} \bibinfo{year}{2025}\natexlab{}.
\newblock \bibinfo{title}{Agentic Retrieval-Augmented Generation: A Survey on Agentic RAG}.
\newblock
\newblock
\showeprint[arxiv]{2501.09136}~[cs.AI]
\urldef\tempurl%
\url{https://arxiv.org/abs/2501.09136}
\showURL{%
\tempurl}


\bibitem[\protect\citeauthoryear{Song, Jiang, Min, Chen, Chen, Zhao, Fang, and Wen}{Song et~al\mbox{.}}{2025}]%
        {r1Searcher}
\bibfield{author}{\bibinfo{person}{Huatong Song}, \bibinfo{person}{Jinhao Jiang}, \bibinfo{person}{Yingqian Min}, \bibinfo{person}{Jie Chen}, \bibinfo{person}{Zhipeng Chen}, \bibinfo{person}{Wayne~Xin Zhao}, \bibinfo{person}{Lei Fang}, {and} \bibinfo{person}{Ji-Rong Wen}.} \bibinfo{year}{2025}\natexlab{}.
\newblock \bibinfo{title}{R1-Searcher: Incentivizing the Search Capability in LLMs via Reinforcement Learning}.
\newblock
\newblock
\showeprint[arxiv]{2503.05592}~[cs.AI]
\urldef\tempurl%
\url{https://arxiv.org/abs/2503.05592}
\showURL{%
\tempurl}


\bibitem[\protect\citeauthoryear{Tao and Wu}{Tao and Wu}{2014}]%
        {SMV}
\bibfield{author}{\bibinfo{person}{Yongquan Tao} {and} \bibinfo{person}{Shengli Wu}.} \bibinfo{year}{2014}\natexlab{}.
\newblock \showarticletitle{Query Performance Prediction By Considering Score Magnitude and Variance Together}. In \bibinfo{booktitle}{\emph{Proceedings of the 23rd ACM International Conference on Conference on Information and Knowledge Management}} (Shanghai, China) \emph{(\bibinfo{series}{CIKM '14})}. \bibinfo{publisher}{Association for Computing Machinery}, \bibinfo{address}{New York, NY, USA}, \bibinfo{pages}{1891–1894}.
\newblock
\showISBNx{9781450325981}
\urldef\tempurl%
\url{https://doi.org/10.1145/2661829.2661906}
\showURL{%
\tempurl}


\bibitem[\protect\citeauthoryear{Tian, Ganguly, and Macdonald}{Tian et~al\mbox{.}}{2025}]%
        {RelevanceAndUtility}
\bibfield{author}{\bibinfo{person}{Fangzheng Tian}, \bibinfo{person}{Debasis Ganguly}, {and} \bibinfo{person}{Craig Macdonald}.} \bibinfo{year}{2025}\natexlab{}.
\newblock \showarticletitle{Is Relevance Propagated from Retriever to Generator in RAG?}. In \bibinfo{booktitle}{\emph{Advances in Information Retrieval: 47th European Conference on Information Retrieval, ECIR 2025, Lucca, Italy}} (Lucca, Italy). \bibinfo{numpages}{18}~pages.
\newblock


\bibitem[\protect\citeauthoryear{Trivedi, Balasubramanian, Khot, and Sabharwal}{Trivedi et~al\mbox{.}}{2023}]%
        {trivedi2023interleaving}
\bibfield{author}{\bibinfo{person}{Harsh Trivedi}, \bibinfo{person}{Niranjan Balasubramanian}, \bibinfo{person}{Tushar Khot}, {and} \bibinfo{person}{Ashish Sabharwal}.} \bibinfo{year}{2023}\natexlab{}.
\newblock \showarticletitle{Interleaving Retrieval with Chain-of-Thought Reasoning for Knowledge-Intensive Multi-Step Questions}. In \bibinfo{booktitle}{\emph{Proceedings of the 61st Annual Meeting of the Association for Computational Linguistics (Volume 1: Long Papers)}}. \bibinfo{pages}{10014--10037}.
\newblock


\bibitem[\protect\citeauthoryear{Varshney, Luo, and Baral}{Varshney et~al\mbox{.}}{2022}]%
        {efficientRAG}
\bibfield{author}{\bibinfo{person}{Neeraj Varshney}, \bibinfo{person}{Man Luo}, {and} \bibinfo{person}{Chitta Baral}.} \bibinfo{year}{2022}\natexlab{}.
\newblock \bibinfo{title}{Can Open-Domain QA Reader Utilize External Knowledge Efficiently like Humans?}
\newblock
\newblock
\showeprint[arxiv]{2211.12707}~[cs.CL]
\urldef\tempurl%
\url{https://arxiv.org/abs/2211.12707}
\showURL{%
\tempurl}


\bibitem[\protect\citeauthoryear{Vlachou}{Vlachou}{2025}]%
        {thesisMariaV}
\bibfield{author}{\bibinfo{person}{Maria Vlachou}.} \bibinfo{year}{2025}\natexlab{}.
\newblock \emph{\bibinfo{title}{Predicting Retrieval Failures in Conversational Recommendation Systems}}.
\newblock Ph.D. thesis. \bibinfo{school}{University of Glasgow}, \bibinfo{address}{Glasgow, UK}.
\newblock


\bibitem[\protect\citeauthoryear{Vlachou and Macdonald}{Vlachou and Macdonald}{2024}]%
        {aPairRatio}
\bibfield{author}{\bibinfo{person}{Maria Vlachou} {and} \bibinfo{person}{Craig Macdonald}.} \bibinfo{year}{2024}\natexlab{}.
\newblock \showarticletitle{Coherence-based Query Performance Measures for Dense Retrieval}. In \bibinfo{booktitle}{\emph{Proceedings of the 2024 ACM SIGIR International Conference on Theory of Information Retrieval}} (Washington DC, USA) \emph{(\bibinfo{series}{ICTIR '24})}. \bibinfo{publisher}{Association for Computing Machinery}, \bibinfo{address}{New York, NY, USA}, \bibinfo{pages}{15–24}.
\newblock
\showISBNx{9798400706813}
\urldef\tempurl%
\url{https://doi.org/10.1145/3664190.3672518}
\showURL{%
\tempurl}


\bibitem[\protect\citeauthoryear{Wang, Yang, Huang, Jiao, Yang, Jiang, Majumder, and Wei}{Wang et~al\mbox{.}}{2022}]%
        {wang2022text}
\bibfield{author}{\bibinfo{person}{Liang Wang}, \bibinfo{person}{Nan Yang}, \bibinfo{person}{Xiaolong Huang}, \bibinfo{person}{Binxing Jiao}, \bibinfo{person}{Linjun Yang}, \bibinfo{person}{Daxin Jiang}, \bibinfo{person}{Rangan Majumder}, {and} \bibinfo{person}{Furu Wei}.} \bibinfo{year}{2022}\natexlab{}.
\newblock \showarticletitle{Text embeddings by weakly-supervised contrastive pre-training}.
\newblock \bibinfo{journal}{\emph{arXiv preprint arXiv:2212.03533}} (\bibinfo{year}{2022}).
\newblock


\bibitem[\protect\citeauthoryear{Wang, Liu, Lin, Li, Ma, and Liang}{Wang et~al\mbox{.}}{2024}]%
        {RAT}
\bibfield{author}{\bibinfo{person}{Zihao Wang}, \bibinfo{person}{Anji Liu}, \bibinfo{person}{Haowei Lin}, \bibinfo{person}{Jiaqi Li}, \bibinfo{person}{Xiaojian Ma}, {and} \bibinfo{person}{Yitao Liang}.} \bibinfo{year}{2024}\natexlab{}.
\newblock \bibinfo{title}{RAT: Retrieval Augmented Thoughts Elicit Context-Aware Reasoning in Long-Horizon Generation}.
\newblock
\newblock
\showeprint{arXiv:2403.05313}


\bibitem[\protect\citeauthoryear{Yang, Yang, Zhang, Hui, Zheng, Yu, Li, Liu, Huang, Wei, Lin, Yang, Tu, Zhang, Yang, Yang, Zhou, Lin, Dang, Lu, Bao, Yang, Yu, Li, Xue, Zhang, Zhu, Men, Lin, Li, Tang, Xia, Ren, Ren, Fan, Su, Zhang, Wan, Liu, Cui, Zhang, and Qiu}{Yang et~al\mbox{.}}{2025}]%
        {qwen25}
\bibfield{author}{\bibinfo{person}{An Yang}, \bibinfo{person}{Baosong Yang}, \bibinfo{person}{Beichen Zhang}, \bibinfo{person}{Binyuan Hui}, \bibinfo{person}{Bo Zheng}, \bibinfo{person}{Bowen Yu}, \bibinfo{person}{Chengyuan Li}, \bibinfo{person}{Dayiheng Liu}, \bibinfo{person}{Fei Huang}, \bibinfo{person}{Haoran Wei}, \bibinfo{person}{Huan Lin}, \bibinfo{person}{Jian Yang}, \bibinfo{person}{Jianhong Tu}, \bibinfo{person}{Jianwei Zhang}, \bibinfo{person}{Jianxin Yang}, \bibinfo{person}{Jiaxi Yang}, \bibinfo{person}{Jingren Zhou}, \bibinfo{person}{Junyang Lin}, \bibinfo{person}{Kai Dang}, \bibinfo{person}{Keming Lu}, \bibinfo{person}{Keqin Bao}, \bibinfo{person}{Kexin Yang}, \bibinfo{person}{Le Yu}, \bibinfo{person}{Mei Li}, \bibinfo{person}{Mingfeng Xue}, \bibinfo{person}{Pei Zhang}, \bibinfo{person}{Qin Zhu}, \bibinfo{person}{Rui Men}, \bibinfo{person}{Runji Lin}, \bibinfo{person}{Tianhao Li}, \bibinfo{person}{Tianyi Tang}, \bibinfo{person}{Tingyu Xia}, \bibinfo{person}{Xingzhang Ren},
  \bibinfo{person}{Xuancheng Ren}, \bibinfo{person}{Yang Fan}, \bibinfo{person}{Yang Su}, \bibinfo{person}{Yichang Zhang}, \bibinfo{person}{Yu Wan}, \bibinfo{person}{Yuqiong Liu}, \bibinfo{person}{Zeyu Cui}, \bibinfo{person}{Zhenru Zhang}, {and} \bibinfo{person}{Zihan Qiu}.} \bibinfo{year}{2025}\natexlab{}.
\newblock \bibinfo{title}{Qwen2.5 Technical Report}.
\newblock
\newblock
\showeprint[arxiv]{2412.15115}~[cs.CL]
\urldef\tempurl%
\url{https://arxiv.org/abs/2412.15115}
\showURL{%
\tempurl}


\bibitem[\protect\citeauthoryear{Yang, Qi, Zhang, Bengio, Cohen, Salakhutdinov, and Manning}{Yang et~al\mbox{.}}{2018}]%
        {hotpotQA}
\bibfield{author}{\bibinfo{person}{Zhilin Yang}, \bibinfo{person}{Peng Qi}, \bibinfo{person}{Saizheng Zhang}, \bibinfo{person}{Yoshua Bengio}, \bibinfo{person}{William Cohen}, \bibinfo{person}{Ruslan Salakhutdinov}, {and} \bibinfo{person}{Christopher~D. Manning}.} \bibinfo{year}{2018}\natexlab{}.
\newblock \showarticletitle{{H}otpot{QA}: A Dataset for Diverse, Explainable Multi-hop Question Answering}. In \bibinfo{booktitle}{\emph{Proceedings of the 2018 Conference on Empirical Methods in Natural Language Processing}}. \bibinfo{publisher}{Association for Computational Linguistics}, \bibinfo{address}{Brussels, Belgium}, \bibinfo{pages}{2369--2380}.
\newblock
\urldef\tempurl%
\url{https://aclanthology.org/D18-1259/}
\showURL{%
\tempurl}


\end{thebibliography}

\appendix
\section{Appendix}~\label{sec:appendix}

In this Appendix, we study one negative example (\ref{ap:negative_eg}) and one positive example (\ref{ap:positive_eg}) output from Search‑R1 with retriever E5. The purpose is to investigate the reasons leading to the degradation of average retrieval quality in later iterations. \ref{ap:insights} presents a brief insight from the analysis of the examples.

\looseness -1 \subsection{Negative Example: Ambiguous Question Causes Drifted Sub-Queries and Low-Quality Answer}~\label{ap:negative_eg}

Figure~\ref{fig:output_example_negative} shows the studied negative example. In this example, the input question ``chris cornell she’s going to change the world'' asks for the name of the Chris Cornell song that contains the lyric ``She’s going to change the world.'' The query is formulated very ambiguously; consequently, it confuses the Agentic RAG model. The correct answer should be ``Can’t Change Me'', but the RAG model instead yields Eric Clapton, a singer known for the song ``Change the World''.

\begin{figure}[h]


\begin{subfigure}{\textwidth}
\begin{tcolorbox}[colback=green!5!white, colframe=green!50!black, title=Output - Iteration 1]

<think>I need to determine if Chris Cornell sang the song ``She's Going to Change the World''. I'll search for it.</think> <search> \textbf{chris cornell She's Going to Change the World} </search>

<information>
\textbf{Doc 1(Title: ``Chris Cornell'')} called ``Chris Cornell''. According to lead singer Shirley Manson, she was obsessed with Cornell at the time of the release of Soundgarden's hit song ``Black Hole Sun''. A demo version of ``Fix Me Now'' has Manson singing Cornell's name. The line ``I'm looking California, and feeling Minnesota'' from Soundgarden's song ``Outshined'' (written by Cornell), inspired the title of Steven Baigelman's 1996 film ``Feeling Minnesota'', starring Keanu Reeves and Cameron Diaz. Cornell's song ``Preaching the End of the World'' from his debut solo album ``Euphoria Morning'', inspired the title of Lorene Scafaria's 2012 film ``Seeking a Friend for the End of

\textbf{Doc 2(Title: ``Chris Cornell'')} the World'', starring Steve Carell and Keira Knightley. In 1985, Cornell started dating Susan Silver, the manager of Alice in Chains, Soundgarden and Screaming Trees, and they got married in 1990. They had a daughter, Lillian Jean, born in June 2000. He and Silver divorced in 2004. In December 2008, Cornell reported via his official website that he had finally won back his collection of 15 guitars after a four-year court battle with Silver. In 2004, he married Vicky Karayiannis, a Paris-based American publicist of Greek heritage. The couple had a daughter together, Toni, in September 2004, and a son,

\textbf{Doc 3(Title: ``Chris Cornell'')} Chris Cornell Christopher John Cornell (né Boyle; July 20, 1964 – May 18, 2017) was an American musician, singer and songwriter. He was best known as the lead vocalist for the rock bands Soundgarden and Audioslave. Cornell was also known for his numerous solo works and soundtrack contributions since 1991, and as the founder and frontman for Temple of the Dog, the one-off tribute band dedicated to his late friend Andrew Wood. Cornell is considered one of the chief architects of the 1990s grunge movement, and is well known for his extensive catalog as a songwriter, his nearly four-octave vocal
</information>

\end{tcolorbox}
\end{subfigure}
\hfill
\begin{subfigure}{\textwidth}
\begin{tcolorbox}[colback=green!5!white, colframe=green!50!black, title=Output - Iteration 2]

<think>I found out that Chris Cornell sang for the rock bands Soundgarden and Audioslave. Now I need to check if he sang the song ``She's Going to Change the World''. I'll search for it.</think> <search> \textbf{She's Going to Change the World} </search>

<information>
\textbf{Doc 1(Title: ``How to Change the World'')} It was very difficult to start a private organization at first, since Hungary was a communist country. So she started with a small group of disabled individuals and taught them simple skills that they could use to work a job, and that grew larger and changed over the years. It eventually became a functional and beautiful community where disabled individuals could come and take up a residency, learn a skill and make a living, and learn how to live independently if possible. There were many times where she failed, or her efforts seemed to be going nowhere. But she never

\textbf{Doc 2(Title: ``How to Change the World'')} method, by stern discipline, by rigid attention to detail, by ceaseless labor, by the fixed determination of an indomitable will.'' She knew she had the ability to make an extremely beneficial impact, and even though it reversed society's norm, she succeeded in fulfilling her dreams. Florence Nightingale will forever be remembered as a great social entrepreneur for her heroic work in the field of nursing. Nightingale's ``indomitable will'' may have been her most instrumental trait in accomplishing her radical ideas and plans. Beginning with strong resistance from her parents, and continuing to the point of defying army officers, her calling

\textbf{Doc 3(Title: ``How to Change the World'')} was a constant struggle against society's standards. Like Nightingale, every social entrepreneur will face opposition at some point in their career. It comes with the very definition of an entrepreneur. Their job is to defy cultural norms to find revolutionary, though often seemingly insane, ideas to solve the problems of life. ``It takes concentrated focus, practical creativity, and a long-term source of energy to advance a system change and to ensure that the change becomes well rooted in institutions and cultures.'' Social entrepreneurs must have the courage and indomitable will of Florence Nightingale to succeed in a world so resistant
</information>

\end{tcolorbox}
\end{subfigure}
\hfill
\end{figure}

\begin{figure}[h]\ContinuedFloat
\begin{subfigure}{\textwidth}
\begin{tcolorbox}[colback=green!5!white, colframe=green!50!black, title=Output - Iteration 3]

<think>I found out that ``She's Going to Change the World'' is a song, but I still need to check if Chris Cornell sang it. I'll search for it.</think> <search> \textbf{who sang She's Going to Change the World} </search>

<information>\textbf{Doc 1(Title: ``Change the World'')} Change the World ``Change the World'' is a song written by Tommy Sims, Gordon Kennedy, and Wayne Kirkpatrick whose best-known version was recorded by the British recording artist Eric Clapton for the soundtrack of the 1996 film, ``Phenomenon''. The track was produced by R\&B record producer Kenneth ``Babyface'' Edmonds. The single release, Clapton recorded for Reprise and Warner Bros. Records, reached the Top 40 in twenty countries and topped the charts in Canada as well as ``Billboard'' magazine's Adult Contemporary and Adult Top 40 charts in the United States. The single was prized with eight awards, among them three Grammy

\textbf{Doc 2(Title: ``Gordon Kennedy (musician)'')} 20 years of friendship, Kennedy remains Frampton's collaborator and performs with him often on tour. As of 2017, Kennedy has composed for or recorded with Bonnie Raitt (five songs, including ``Gypsy in Me''), Reba McEntire, Tim McGraw, Wynonna, Charlie Daniels, Trisha Yearwood, George Strait, Faith Hill (``It Will Be Me''), Carrie Underwood (``The More Boys I Meet''), and others. The enduring worldwide hit song, ``Change the World'', earned Kennedy a Grammy for ``Song of the Year'' in 1996. The song was first recorded in 1995 by Wynonna Judd and later Eric Clapton through the release of the John Travolta film,

\textbf{Doc 3(Title: ``Change the World'')} the line ``If I could change the world...'', leading to the second link, which is repeated three times, before the song's outro, which is identical to the song's intro and interlude. Before Clapton's hit version was released, the song was recorded by country artist Wynonna Judd for her February 1996 album, ``Revelations'' (MCA Records). Her three-minute, nineteen-second take on the song is more of a neo-traditionalist country music track. Judd did not release her version as a single. Judd's interpretation was also released on the compilation album ``Best of America, Volume 2'' for Curb Records on June 3, 2003. Eric
</information>

\end{tcolorbox}
\end{subfigure}
\hfill
\begin{subfigure}{\textwidth}
\begin{tcolorbox}[colback=green!5!white, colframe=green!50!black, title=Output - Answer] 

<think>I found out that the song ``She's Going to Change the World'' was sung by Eric Clapton. Now I can provide the answer.</think> <answer> \textbf{Eric Clapton} </answer>

\end{tcolorbox}
\end{subfigure}
\caption{\looseness -1 The output from Search-R1 for the negative example ``chris cornell she's going to change the world''. \newcontent{In this example, the intermediate queries gradually drift away from the original information need, eventually leading to the wrong answer.}}
\label{fig:output_example_negative}
\end{figure}

To answer this question, Search‑R1 invokes the retriever (E5) three times. In the first retrieval, the intermediate query is identical to the input. From the reasoning preceding the search, we observe that the LLM treats ``She's going to change the world'' as a song title rather than a lyric excerpt. In the second retrieval, the intermediate query omits ``Chris Cornell'', one of the two key entities from the original question. As a result, the topic drifts significantly. In the third search, the retrieved documents focus on a song titled ``Changed the World'', which is only weakly related to the intermediate query, indicating a lower-quality retrieval.

The QPP estimates (using Max(Score)) for the three iterations are $0.8370$, $0.7870$ and $0.8511$, which are lower than those in the positive examples discussed in the next section (Max(Score)$>0.9$). It is worth noting that retrieval quality does not necessarily decline monotonically. In this case, the ambiguity in the original question causes the LLM to misinterpret the user's information need, ultimately leading to an incorrect answer.

\subsection{Positive Example: Specific Question with High Quality Retrievals}~\label{ap:positive_eg}

Figure~\ref{fig:output_example_positive} shows the positive example under study. In this case, the input question: ``when was the last time villanova won the ncaa basketball championship?'' represents a very clear information need. The Search‑R1 model uses two reasoning-retrieval iterations to answer this question. After the first retrieval, which reformulates the original query, the model initiates a second search — another reformulation of the same question. After two rounds, the model correctly outputs 2018 as the answer.

The QPP estimates (using Max(Score)) for these two iterations are $0.9263$ and $0.9268$, respectively. Compared to the negative example with three iterations, these high values indicate strong retrieval quality. Such high-quality retrievals enable the RAG model to arrive at the correct answer. Since questions requiring fewer iterations tend to rely on more effective intermediate retrievals, their average estimated query performance is generally higher in earlier iterations.

However, this example also reveals a persistent issue in Search‑R1 and other agentic models: repetition of intermediate queries. With E5 as the retriever and a cutoff of 3 iterations, $262$ out of $3610$ queries generated identical sub-queries during reasoning. This indicates insufficient utilisation of retrieved documents in current Agentic models. Incorporating QPP or other signals into the LLM could help achieve more efficient use of retrieved information.

\begin{figure}[h]


\begin{subfigure}{\textwidth}
\begin{tcolorbox}[colback=green!5!white, colframe=green!50!black, title=Output - Iteration 1]

<think>I need to find the last time Villanova won the NCAA basketball championship. I'll search for it.</think> <search> \textbf{when was the last time Villanova won the NCAA basketball championship} </search>

<information>\textbf{Doc 1(Title: ``Villanova Wildcats men's basketball'')} a team for the game. Their 2016 NCAA Championship, is referred to as ``The Perfect Ending'' and is the only NCAA Men's Championship game to be won on a buzzer beater, as Kris Jenkins drained a shot as time expired. They made the Final Four in 1939, 1971, 1985, 2009, 2016 and 2018; their six Final Four appearances are 13th most all-time. As of 2018, they have an NCAA Tournament record of 64–36 (). Villanova has defeated six No. 1 seeds in the NCAA tournament (Michigan and Georgetown in 1985, Pittsburgh in 2009, and Kansas and North Carolina in 2016,

\textbf{Doc 2(Title: ``Villanova Wildcats men's basketball'')} and Kansas in 2018), which is sixth most all-time. The Villanova Wildcats have appeared in the NCAA Tournament 37 times, the eighth highest total in NCAA history. They have won the Big East regular season championship eight times, most recently winning four straight from 2014 to 2017. They won the Big East Tournament in 1995, 2015, 2017, and 2018. Villanova entered the 2016–2017 season with an all-time winning percentage of (.648), placing the Wildcats tied for 13th among all NCAA Division I basketball programs. Through 2018, Villanova has 1,753 wins, which is 23rd among Division I men's basketball teams. Villanova

\textbf{Doc 3(Title: ``Villanova Wildcats men's basketball'')} and were the national champions in 1985, 2016, and 2018. Villanova has won as the underdog (based on Tournament seeding) 16 times, more than any other program. Villanova is one of only two programs (the other being Ohio State) that has played in the NCAA Tournament in every decade since the 1930s. ``The NCAA began seeding the tournament with the 1979 edition.'' The Wildcats have appeared in the National Invitation Tournament (NIT) 17 times. Their combined record is 24–17. They were NIT Champions in 1994. The Wildcats appeared in the only National Campus Basketball Tournament. Their record is 0–1.
</information>

\end{tcolorbox}
\end{subfigure}
\hfill
\begin{subfigure}{\textwidth}
\begin{tcolorbox}[colback=green!5!white, colframe=green!50!black, title=Output - Iteration 2]

<think>I found out that Villanova won the NCAA basketball championship in 1985, 2016, and 2018. Now I need to find the last time they won the championship.</think> <search> \textbf{last time Villanova won the NCAA basketball championship} </search>

<information>\textbf{Doc 1(Title: ``Villanova Wildcats men's basketball'')} a team for the game. Their 2016 NCAA Championship, is referred to as ``The Perfect Ending'' and is the only NCAA Men's Championship game to be won on a buzzer beater, as Kris Jenkins drained a shot as time expired. They made the Final Four in 1939, 1971, 1985, 2009, 2016 and 2018; their six Final Four appearances are 13th most all-time. As of 2018, they have an NCAA Tournament record of 64–36 (). Villanova has defeated six No. 1 seeds in the NCAA tournament (Michigan and Georgetown in 1985, Pittsburgh in 2009, and Kansas and North Carolina in 2016,

\textbf{Doc 2(Title: ``Villanova Wildcats men's basketball'')} and Kansas in 2018), which is sixth most all-time. The Villanova Wildcats have appeared in the NCAA Tournament 37 times, the eighth highest total in NCAA history. They have won the Big East regular season championship eight times, most recently winning four straight from 2014 to 2017. They won the Big East Tournament in 1995, 2015, 2017, and 2018. Villanova entered the 2016–2017 season with an all-time winning percentage of (.648), placing the Wildcats tied for 13th among all NCAA Division I basketball programs. Through 2018, Villanova has 1,753 wins, which is 23rd among Division I men's basketball teams. Villanova

\textbf{Doc 3(Title: ``Villanova Wildcats men's basketball'')} and were the national champions in 1985, 2016, and 2018. Villanova has won as the underdog (based on Tournament seeding) 16 times, more than any other program. Villanova is one of only two programs (the other being Ohio State) that has played in the NCAA Tournament in every decade since the 1930s. ``The NCAA began seeding the tournament with the 1979 edition.'' The Wildcats have appeared in the National Invitation Tournament (NIT) 17 times. Their combined record is 24–17. They were NIT Champions in 1994. The Wildcats appeared in the only National Campus Basketball Tournament. Their record is 0–1.
</information>

\end{tcolorbox}
\end{subfigure}

\begin{subfigure}{\textwidth}
\begin{tcolorbox}[colback=green!5!white, colframe=green!50!black, title=Output - Answer] 

<think>I found out that Villanova won the NCAA basketball championship in 1985, 2016, and 2018. The last time Villanova won the championship was in 2018.</think> <answer> \textbf{2018} </answer>

\end{tcolorbox}
\end{subfigure}
\caption{\looseness -1 The output from Search-R1 for the positive example ``when was the last time villanova won the ncaa basketball championship''. \newcontent{In this example, through multiple retrievals based on reformulations of the original query, Search-R1 model reaches the correct answer.}}
\label{fig:output_example_positive}
\end{figure}


\subsection{Insights from the Case Study}~\label{ap:insights}

\newcontent{From the above analysis, we observe that questions requiring more iterations in their reasoning process tend to involve lower-quality intermediate retrievals, which do not positively contribute to the correct final answer. Conversely, reasoning trajectories that lead to correct answers generally involve fewer iterations, generally yielding higher-quality retrievals. When we compute the average (predicted) retrieval quality by iteration, the early iterations — populated by questions with shorter reasoning chains - exhibit higher mean quality than later iterations.}

\newcontent{However, in the positive example (see Appendix \ref{ap:positive_eg}), we still observe the current Agentic RAG models repeating identical queries in intermediate retrieval steps. This redundancy can reduce the efficiency of the system and underscores the need to improve the Agentic workflow further.}
\end{document}